\documentclass[11pt,nofootinbib,showpacs]{revtex4}
\usepackage{amsfonts,amssymb,amsmath,graphicx,multirow,dsfont}
\def\dac{\displaystyle\frac}

\def\[{\left[}
\def\]{\right]}
\def\({\left(}
\def\){\right)}

\def\toto{\leftrightarrow}
\def\ot{\leftarrow}

\newcommand{\diag}{\mathop{\rm diag}\nolimits}

\begin{document}

\baselineskip7mm

\title{Realistic compactification in spatially flat vacuum cosmological models in cubic Lovelock gravity: High-dimensional case}

\author{Sergey A. Pavluchenko}
\affiliation{Programa de P\'os-Gradua\c{c}\~ao em F\'isica, Universidade Federal do Maranh\~ao (UFMA), 65085-580, S\~ao Lu\'is, Maranh\~ao, Brazil}

\begin{abstract}
In this paper we perform systematic investigation of all possible regimes in spatially flat vacuum cosmological models in cubic Lovelock gravity. The spatial section is considered
as a product of three- and extra-dimensional isotropic subspaces, and the former represents our Universe. As the equations of motion are different for $D=3, 4, 5$ and
general $D \geqslant 6$ cases, we considered them all separately. This is the second paper of the series, and we consider $D=5$ and general $D \geqslant 6$ cases here.
For each $D$ case we found critical values for $\alpha$ (Gauss-Bonnet coupling) and $\beta$ (cubic Lovelock coupling)
which separate different dynamical cases, isotropic and
anisotropic exponential solutions, and study the dynamics in each region to find all regimes for all initial conditions and for arbitrary values of $\alpha$ and $\beta$.
The results suggest that in all $D \geqslant 3$ there are regimes with realistic compactification originating from so-called ``generalized Taub'' solution. The endpoint of the
compactification regimes is either anisotropic exponential solution (for $\alpha > 0$, $\mu \equiv \beta/\alpha^2 < \mu_1$ (including entire $\beta < 0$)) or standard Kasner regime
(for $\alpha > 0$, $\mu > \mu_1$).
For $D \geqslant 8$ there is additional regime which originates from high-energy (cubic Lovelock) Kasner regime and ends as anisotropic exponential solution.
It exists in two domains: $\alpha > 0$, $\beta < 0$, $\mu \leqslant \mu_4$ and entire $\alpha > 0$, $\beta > 0$. Let us note that for
$D \geqslant 8$ and $\alpha > 0$, $\beta < 0$, $\mu < \mu_4$ there are two realistic compactification regimes which exist at the same time and have two different anisotropic exponential solutions as a
future asymptotes. For $D \geqslant 8$ and $\alpha > 0$,
$\beta > 0$, $\mu < \mu_2$ there are two realistic compactification regimes but they lead to the same anisotropic exponential solution. This behavior is quite different
from the Einstein-Gauss-Bonnet case.
There are two more unexpected observations among the results -- all realistic compactification regimes exist only for $\alpha > 0$ and there is no smooth transition from high-energy Kasner regime
to the low-energy regime with realistic compactification.
\end{abstract}

\pacs{04.50.-h, 11.25.Mj, 98.80.Cq}






\maketitle

\section{Introduction}

It is not widely known but the idea of extra dimensions is older then General Relativity (GR) itself. Indeed, the first known extra-dimensional model was introduced by
Nordstr\"om in 1914~\cite{Nord1914} -- it unified Nordstr\"om's second gravity theory~\cite{Nord_2grav} with Maxwell's electromagnetism. Soon after that Einstein proposed GR~\cite{einst}, but
it took years before it was accepted: during the solar eclipse of 1919, the
bending of light near the Sun was measured and the
deflection angle was in perfect agreement with GR, while Nordstr\"om's theory predicted a zeroth deflection angle, as most of the scalar gravity theories do.

But the idea of extra dimensions was not forgotten -- in 1919 Kaluza proposed~\cite{KK1} a very similar model but based on GR: in his model five-dimensional Einstein equations are  decomposed into $4D$ Einstein equations and Maxwell's electromagnetism. But for such decomposition to exist, the extra dimensions should be ``curled'' or compactified into a circle and ``cylindrical conditions'' should be imposed. The work by Kaluza was followed by Klein who proposed~\cite{KK2, KK3} a nice quantum mechanical interpretation of this extra dimension and so the theory, called Kaluza-Klein after its founders, was finalized. It is interesting to note that their theory unified all
known interactions at their time. With flow of time, more interactions were discovered and it became clear that to unify all of them, more extra dimensions are needed. At present, one of the promising theories to unify all interactions is M/string theory.

One of the distinguishing features of M/string theories is the presence in the Lagrangian of the corrections which are quadratic in curvature.
Scherk and Schwarz~\cite{sch-sch}  demonstrated that $R^2$ and
$R_{\mu \nu} R^{\mu \nu}$ terms are presented in the Lagrangian of the Virasoro-Shapiro
model~\cite{VSh1, VSh2}; presence of the term of $R^{\mu \nu \lambda \rho}
R_{\mu \nu \lambda \rho}$ type was found in~\cite{Candelas_etal} for the low-energy limit
of the $E_8 \times E_8$ heterotic superstring theory~\cite{Gross_etal} to match the kinetic term
of the Yang-Mills field. Later Zwiebach demonstrated~\cite{zwiebach} that the only
combination of quadratic terms that leads to a ghost-free nontrivial gravitation
interaction is the Gauss-Bonnet (GB) term:

$$
L_{GB} = L_2 = R_{\mu \nu \lambda \rho} R^{\mu \nu \lambda \rho} - 4 R_{\mu \nu} R^{\mu \nu} + R^2.
$$

\noindent This term was first discovered
by Lanczos~\cite{Lanczos1, Lanczos2} (and so sometimes it is referred to
as the Lanczos term), is an Euler topological invariant in (3+1)-dimensional
space-time, but in (4+1) and higher dimensions it gives nontrivial contribution to the equations of motion.
Zumino~\cite{zumino} extended Zwiebach's result on the
higher-order curvature terms, supporting the idea that the low-energy limit of the unified
theory might have a Lagrangian density as a sum of contributions of different powers of curvature. The sum of all possible Euler topological invariants, which give nontrivial
contribution to the equations of motion in a particular number of space-time dimensions, form  more general Lovelock gravity~\cite{Lovelock}.

When someone mention extra spatial dimensions, the natural question arises -- where they are? Our everyday experience clearly points on three spatial dimensions, and
experiments in physics and theory support this (for example, in Newtonian gravity if there are more then three space dimensions, no stable orbits exist, while we clearly see they do).
The string theorists working with extra dimensions proposed an answer -- the extra spatial dimensions are compact -- they are compactified on a very small scale, so small that we
cannot sense them with our level of equipment. But with an answer like that, another natural question comes to mind -- how come that they are compact? The answer to this question is not that simple. One of
the ways to hide extra dimensions and to recover four-dimensional physics, is so-called ``spontaneous compactification''. Exact static solutions of this type with the metric as a cross product of a
(3+1)-dimensional Minkowski space-time and a constant curvature ``inner space'',  were found for the first time in~\cite{add_1} (the generalization for
a constant curvature Lorentzian manifold was done in~\cite{Deruelle2}).
For cosmology, it is more useful to consider the four-dimensional space-time given by a Friedmann-Robertson-Walker metric, and the size of extra dimensions to be time-dependent rather
then static. In~\cite{add_4} it was demonstrated that to have a more realistic model it is necessary to consider the dynamical evolution of the extra-dimensional scale factor.
In~\cite{Deruelle2}, the equations of motion with time-dependent scale factors were written for arbitrary Lovelock order in the special case of a spatially flat metric (the results were further proven and extended in~\cite{prd09}).
The results of~\cite{Deruelle2} were further analyzed for the special case of 10 space-time dimensions in~\cite{add_10}.
In~\cite{add_8}, the dynamical compactification was studied with use of the Hamiltonian formalism.
More recently, studies of the spontaneous  compactifications were made in~\cite{add13}, where
the dynamical compactification of the (5+1) Einstein-Gauss-Bonnet (EGB) model was considered; in \cite{MO04, MO14} with different metric {\it Ans\"atze} for scale factors
corresponding to (3+1)- and extra-dimensional parts. Also, apart from the
cosmology, the recent analysis was focused on
properties of black holes in Gauss-Bonnet~\cite{alpha_12, add_rec_1, add_rec_2, addn_1, addn_2, add_o_1, add_o_2, add_o_3} and Lovelock~\cite{add_rec_3, add_rec_4, addn_3, addn_4, addn_4.1} gravities, features of gravitational collapse in these
theories~\cite{addn_5, addn_6, addn_7}, general features of spherical-symmetric solutions~\cite{addn_8}, and many others.

If we want to find exact cosmological solutions, the most common {\it Ansatz} for the scale factor is either exponential or power-law.
Exact solutions with exponents  for both  the (3+1)- and extra-dimensional scale factors were studied for the first time in~\cite{Is86}, and exponentially expanding (3+1)-dimensional
part and an exponentially shrinking extra-dimensional scale factor were described.
Power-law solutions have been considered  in~\cite{Deruelle1, Deruelle2} and more  recently in~\cite{mpla09, prd09, Ivashchuk, prd10, grg10} so that by now there is more-or-less complete description of the solutions of this kind
(see also~\cite{PT} for comments regarding different physical branches of the power-law solutions).
Solutions with exponential scale factors~\cite{KPT} have also been studied in detail, namely, the models with both variable~\cite{CPT1} and constant~\cite{CST2} volume; the general scheme for
finding anisotropic exponential solutions in EGB gravity was developed and generalized for general Lovelock gravity of any order and in any dimensions~\cite{CPT3}. The stability of the exponential solutions was addressed in~\cite{my15}
(see also~\cite{iv16} for stability of general exponential
solutions in EGB gravity), and it was
demonstrated that only a handful of the solutions found and described in~\cite{CPT3} could be called ``stable'', while the most of them are either unstable or have neutral/marginal stability.

In order to find all possible cosmological regimes in EGB gravity, one needs to go beyond an exponential or power-law {\it Ansatz} and keep the scale factor generic.
We are especially interested in models that allow dynamical compactification, so that we
consider the metric to be the product of a spatially three-dimensional and extra-dimensional parts. In that case the three-dimensional part is ``our Universe'' and we expect for this part to expand
while the extra-dimensional part should be suppressed in size with respect to the three-dimensional one. In~\cite{CGP1} we demonstrated the there exist the phenomenologically
sensible regime when the curvature of the extra dimensions is negative and the EGB theory does not admit a maximally-symmetric solution. In this case both the
three-dimensional Hubble parameter and the extra-dimensional scale factor asymptotically tend to the constant values. In~\cite{CGP2} we continued investigation of this case and performed a detailed analysis of the cosmological dynamics in this model
with generic couplings. Recent analysis of this model~\cite{CGPT} revealed that, with an additional constraint on couplings, Friedmann-type late-time behavior
could be restored.

With the exponential and power-law solutions described in the mentioned above papers, another natural question arise -- could these solutions describe realistic compactification
(i.e., with proper both past and future asymptotes) or
are they just solutions with no connection to the reality? To answer this question, we have considered the cosmological model in EGB gravity with the spatial part being the product
of three- and extra dimensional parts with both subspaces being spatially flat. As both subspaces are spatially flat, we can rewrite the equations of motion in terms of Hubble
parameters, so that they become first order differential equations and could be analytically analyzed to find all possible regimes, asymptotes, exponential and power-law solutions.
For vacuum EGB model it was done in~\cite{my16a} and reanalyzed in~\cite{my18a}. The results suggest that the vacuum model has two physically viable regimes -- first of them is the smooth transition from high-energy GB Kasner (or generalized Taub solution) to low-energy GR Kasner. This regime
exists for $\alpha > 0$ (Gauss-Bonnet coupling) at $D=1,\,2$ (the number of extra dimensions) and for $\alpha < 0$ at $D \geqslant 2$ (so that at $D=2$ it appears for both signs of $\alpha$). Another viable regime is the smooth
transition from high-energy GB
Kasner to anisotropic exponential solution with expanding three-dimensional section (``our Universe'') and contracting extra dimensions; this regime occurs only for $\alpha > 0$ and at $D \geqslant 2$.

The same analysis but for EGB model with $\Lambda$-term was performed in~\cite{my16b, my17a} and reanalyzed in~\cite{my18a}. The results suggest that the only realistic regime is the transition from high-energy GB Kasner to anisotropic exponential
solution, it requires $D \geqslant 2$, see~\cite{my16b, my17a, my18a} for exact limits on ($\alpha, \Lambda$) in each particular $D$.
The low-energy GR Kasner is
forbidden in the presence of the $\Lambda$-term so the corresponding transition does not occur.

In these studies we have made two important assumptions -- we considered both subspaces being isotropic and spatially flat. But what will happens in we lift these conditions?
Indeed, the spatial section
being a product of two isotropic spatially-flat subspaces could hardly be called ``natural'', so that we considered the effects of anisotropy and spatial curvature in~\cite{PT2017}. The initial anisotropy affects the results greatly -- indeed, say, in vacuum $(4+1)$-dimensional EGB gravity with Bianchi-I-type metric (all the directions are independent) the only future asymptote is nonstandard singularity~\cite{prd10}. Our analysis~\cite{PT2017} suggest that the transition from Gauss-Bonnet Kasner
regime to anisotropic exponential expansion (with expanding
three and contracting extra dimensions) is stable with respect to breaking the symmetry within both three- and extra-dimensional subspaces. However, the details of the dynamics in
$D=2$ and $D \geqslant 3$ are different -- in the latter there exist anisotropic exponential solutions with ``wrong'' spatial splitting and all of them are accessible from generic
initial conditions. For instance, in $(6+1)$-dimensional space-time there are anisotropic exponential solutions with $[3+3]$ and $[4+2]$ spatial splittings, and some of the initial
conditions in the vicinity of $E_{3+3}$ actually end up in $E_{4+2}$ -- the exponential solution with four and two isotropic subspaces. In other words, generic initial conditions
could easily end up with ``wrong'' compactification, giving ``wrong'' number of expanding spatial dimensions (see~\cite{PT2017} for details).

The effect of the spatial curvature on the cosmological dynamics also could be dramatic -- say, positive curvature changes the inflationary asymptotic~\cite{infl1, infl2}. In the case
of EGB gravity the influence of the spatial curvature
reveal itself only if the curvature of the extra dimensions is negative and $D \geqslant 3$
-- in that case there exist  ``geometric frustration'' regime, described in~\cite{CGP1, CGP2} and further investigated in~\cite{CGPT}. Full investigation of the spatial curvature
effects on the existing regimes could be found in~\cite{PT2017}.

The current manuscript is a direct continuation of~\cite{my18b}, where the same analysis was performed for low-$D$ ($D=3, 4$) cases. It also
could be called a spiritual successor of~\cite{my16a, my16b, my17a, my18a} -- now we are performing the same analysis but for cubic Lovelock gravity. In this
paper we consider only vacuum case, $\Lambda$-term case, as well as possible influence of anisotropy, spatial curvature and different kinds of matter source are to be considered in
the papers to follow.

The manuscript is structured as follows: first we introduce Lovelock gravity and derive the equations of motion in the general form for the spatially-flat (Bianchi-I-type) metrics. Then we add our {\it Ansatz} and write down
simplified equations. After that we describe the scheme we are going to use and give some important comments on the existing regimes. After that we consider particular cases with $D=5$ and general $D \geqslant 6$ number of extra dimensions.
In each section, dedicated to the particular case, we describe it and briefly summarize its features. Finally we summarize all the cases,  discuss their differences and similarities, and describe the effect of $D$ on
the features. After that we compare the dynamics in this cubic Lovelock with the dynamics in quadratic Lovelock (Einstein-Gauss-Bonnet) case, described in~\cite{my16a, my18a}. At last, we draw conclusions and formulate
perspective directions for further investigations.

\section{Equations of motion}

Lovelock gravity~\cite{Lovelock} has the following structure: its Lagrangian is constructed from terms

\begin{equation}
L_n = \frac{1}{2^n}\delta^{i_1 i_2 \dots i_{2n}}_{j_1 j_2 \dots
j_{2n}} R^{j_1 j_2}_{i_1 i_2}
 \dots R^{j_{2n-1} j_{2n}}_{i_{2n-1} i_{2n}}, \label{lov_lagr}
\end{equation}

\noindent where $\delta^{i_1 i_2 \dots i_{2n}}_{j_1 j_2 \dots
j_{2n}}$ is the generalized Kronecker delta of the order $2n$.
One can verify that $L_n$ is Euler invariant in $\mathcal{D} < 2n$ spatial dimensions and so it does not give nontrivial contribution into the equations of motion. Then the
Lagrangian density for any given $\mathcal{D}$ spatial dimensions is sum of all Lovelock invariants (\ref{lov_lagr}) upto $n=\[\dac{\mathcal{D}}{2}\]$ which give nontrivial contributions
into equations of motion:

\begin{equation}
{\cal L}= \sqrt{-g} \sum_n c_n L_n, \label{lagr}
\end{equation}

\noindent where $g$ is the determinant of metric tensor,
$c_n$ are coupling constants of the order of Planck length in $\mathcal{D}$
dimensions and we assume summation over all $n$ in consideration. The metric {\it ansatz} has the form

\begin{equation}\label{metric}
g_{\mu\nu} = \diag\{ -1, a_1^2(t), a_2^2(t),\ldots, a_n^2(t)\}.
\end{equation}

\noindent As we mentioned, we are interested in the dynamics in cubic Lovelock gravity, so we consider $n$ up to three ($n=0$ is the boundary term while $n=1$ is Einstein-Hilbert, $n=2$ is Gauss-Bonnet and $n=3$ is cubic
Lovelock contributions).
Substituting metric (\ref{metric}) into the Lagrangian and following the standard procedure gives us the equations of motion:

\begin{equation}
\begin{array}{l}
2 \[ \sum\limits_{j\ne i} (\dot H_j + H_j^2)
+ \sum\limits_{\substack{\{ k > l\} \\ \ne i}} H_k H_l \] + 8\alpha \[ \sum\limits_{j\ne i} (\dot H_j + H_j^2) \sum\limits_{\substack{\{k>l\} \\ \ne \{i, j\}}} H_k H_l +
3 \sum\limits_{\substack{\{ k > l >  \\   m > n\} \ne i}} H_k H_l H_m H_n \] + \\ \\
+ 144\beta\[ \sum\limits_{j\ne i} (\dot H_j + H_j^2) \sum\limits_{\substack{\{k>l>m> \\ n\} \ne \{i, j\}}} H_k H_l H_m H_n + 5 \sum\limits_{\substack{\{ k > l > m >  \\   n > p > q\} \ne i}} H_k H_l H_m H_n H_p H_q   \]
- \Lambda = 0
\end{array} \label{dyn_gen}
\end{equation}

\noindent as the $i$th dynamical equation. The first Lovelock term---the Einstein-Hilbert contribution---is in the first set of brackets, the second term---Gauss-Bonnet---is in the second set and the third -- cubic Lovelock
term---is in the third set; $\alpha$
is the coupling constant for the Gauss-Bonnet contribution while $\beta$ is the coupling constant for cubic Lovelock; we put the corresponding constant for Einstein-Hilbert contribution to unity\footnote{So that effectively $\alpha$ is a ratio of Gauss-Bonnet coupling to the Einstein-Hilbert one, similarly $\beta$ is a ratio of cubic Lovelock coupling to the Einstein-Hilbert one.}.
Since we consider spatially flat cosmological models, scale
factors do not hold much physical sense and the equations are rewritten in terms of the Hubble parameters $H_i = \dot a_i(t)/a_i(t)$. Apart from the dynamical equations, we write down the constraint equation

\begin{equation}
\begin{array}{l}
2 \sum\limits_{i > j} H_i H_j + 24\alpha \sum\limits_{\substack{i > j >\\  k > l}} H_i H_j H_k H_l + 720\beta \sum\limits_{\substack{i > j > k \\ > l> m > n}} H_i H_j H_k H_l H_m H_n= \Lambda.
\end{array} \label{con_gen}
\end{equation}

As we mentioned in the Introduction,
we want to investigate the particular case with the scale factors split into two parts -- separately three dimensions (three-dimensional isotropic subspace), which are supposed to represent our Universe, and the remaining represent the extra dimensions ($D$-dimensional isotropic subspace). So we put $H_1 = H_2 = H_3 = H$ and $H_4 = \ldots = H_{D+3} = h$ ($D$ designs the number of additional dimensions) and the
equations take the following form: the
dynamical equation that corresponds to $H$,

\begin{equation}
\begin{array}{l}
2 \[ 2 \dot H + 3H^2 + D\dot h + \dac{D(D+1)}{2} h^2 + 2DHh\] + 8\alpha \[ 2\dot H \(DHh + \dac{D(D-1)}{2}h^2 \) + \right. \\
 \\ \left. + D\dot h \(H^2 + 2(D-1)Hh + \dac{(D-1)(D-2)}{2}h^2 \) +
2DH^3h + \dac{D(5D-3)}{2} H^2h^2 + \right. \\
\\ \left. + D^2(D-1) Hh^3 + \dac{(D+1)D(D-1)(D-2)}{8} h^4 \] +  \\ \\
+  144\beta \[ \dot H \(Hh^3 \dac{D(D-1)(D-2)}{3} +
h^4\dac{D(D-1)(D-2)(D-3)}{12} \) + \right. \\ \\
+ \left. D \dot h
\(H^2 h^2 \dac{(D-1)(D-2)}{2} +  Hh^3 \dac{(D-1)(D-2)(D-3)}{3} + \right. \right. \\
\\ + \left. \left.  h^4\dac{(D-1)(D-2)(D-3)(D-4)}{24} \) +  H^3h^3 \dac{D(D-1)(D-2)}{3} + \right. \\ \\
+ \left. H^2h^4 \dac{D(D-1)(D-2)(7D-9)}{24} +  Hh^5 \dac{D^2(D-1)(D-2)(D-3)}{12} + \right. \\ \\
+ \left. h^6 \dac{(D+1)D(D-1)(D-2)(D-3)(D-4)}{144} \]
 - \Lambda=0,
\end{array} \label{H_gen}
\end{equation}

\noindent the dynamical equation that corresponds to $h$,

\begin{equation}
\begin{array}{l}
2 \[ 3 \dot H + 6H^2 + (D-1)\dot h + \dac{D(D-1)}{2} h^2 + 3(D-1)Hh\] + 8\alpha \[ 3\dot H \(H^2 + \right. \right. \\
\\ \left. \left. + 2(D-1)Hh +  \dac{(D-1)(D-2)}{2}h^2 \) +  (D-1)\dot h \(3H^2 + 3(D-2)Hh + \right. \right. \\
\\  \left. \left. +
\dac{(D-2)(D-3)}{2}h^2 \) + 3H^4 +  9(D-1)H^3h + 3(D-1)(2D-3) H^2h^2 + \right. \\
\\ \left. + \dac{3(D-1)^2 (D-2)}{2} Hh^3 +   \dac{D(D-1)(D-2)(D-3)}{8} h^4 \] + \\ \\  + 144\beta\[ \dot H \( H^2 h^2 \dac{3(D-1)(D-2)}{2} +  Hh^3(D-1)(D-2)(D-3) + \right. \right. \\
\\ + \left. \left.
h^4 \dac{(D-1)(D-2)(D-3)(D-4)}{8} \)  + (D-1)\dot h \(  H^3 h (D-2) + \right. \right. \\
\\ + \left. \left. H^2h^2 \dac{3(D-2)(D-3)}{2} +  Hh^3 \dac{(D-2)(D-3)(D-4)}{2} + \right. \right. \\
\\  \left. \left. +
h^4 \dac{(D-2)(D-3)(D-4)(D-5)}{24}  \) + H^4 h^2 \dac{3(D-1)(D-2)}{2} + \right. \\ \\ \left. + H^3h^3 \dac{(D-1)(D-2)(11D-27)}{6} +  H^2h^4 \dac{3(D-1)(D-2)^2(D-3)}{4} +
\right. \\ \\ \left. + Hh^5 \dac{(D+1)(D-1)(D-2)(D-3)(D-4)}{12} + \right. \\ \\ \left. +
 h^6\dac{D(D-1)(D-2)(D-3)(D-4)(D-5)}{144} \]
- \Lambda =0,
\end{array} \label{h_gen}
\end{equation}

\noindent and the constraint equation,

\begin{equation}
\begin{array}{l}
2 \[ 3H^2 + 3DHh + \dac{D(D-1)}{2} h^2 \] + 24\alpha \[ DH^3h + \dac{3D(D-1)}{2}H^2h^2 + \right. \\ \\ \left. + \dac{D(D-1)(D-2)}{2}Hh^3 +  \dac{D(D-1)(D-2)(D-3)}{24}h^4\] +
 720\beta \[ H^3 h^3 \dac{D(D-1)(D-2)}{6} + \right. \\ \\ \left. +
 H^2 h^4 \dac{D(D-1)(D-2)(D-3)}{8} +  Hh^5 \dac{D(D-1)(D-2)(D-3)(D-4)}{40} + \right. \\ \\ \left. + h^6 \dac{D(D-1)(D-2)(D-3)(D-4)(D-5)}{720}    \] = \Lambda.
\end{array} \label{con2_gen}
\end{equation}

Looking at (\ref{H_gen})--(\ref{con2_gen}) one can notice that the structure of the equations depends on the number of extra dimensions $D$ (terms with $(D-4)$ multiplier nullifies in $D=4$ and so on).
In previous papers, dedicated to study cosmological dynamics in EGB gravity, we
performed analysis in all dimensions, sensitive to EGB case~\cite{my16a, my16b, my17a, my18a}. In the cubic Lovelock, the structure of the equations of motion is different in $D=3, 4, 5$ and in the general $D \geqslant 6$ cases. In the directly previous paper~\cite{my18b} we performed the analysis for $D=3$ and $D=4$ cases, leaving $D=5$ and the general
$D \geqslant 6$ cases for this paper.
Also, similar to~\cite{my18b}, since the these papers are dedicated to the vacuum case, we have $\Lambda \equiv 0$.

\section{General scheme}

The procedure of the analysis is exactly the same as described in our previous papers~\cite{my16a, my16b, my17a, my18a}. In the directly previous paper~\cite{my18b} this scheme is
described in great detail for $D=3$ case. In the current paper we just list the scheme and use it with almost no details. So the scheme  is as follows:

\begin{itemize}

 \item we solve (\ref{con2_gen}) with respect to $H$ -- one can see that it is cubic with respect to $H$ and sixth order with respect to $h$, so that to have analytical solutions, we solve it for $H$; and
 as a result we have three branches $H_1$, $H_2$ and $H_3$. In lower-dimensional cases we wrote down solutions explicitly, but in higher dimensions they become quite bulky, so we draw $H(h)$ curves instead. If we take the discriminant of (\ref{con2_gen}) with respect to $H$, and then its discriminant with respect to $h$, we obtain critical values for $(\alpha, \beta)$ which separate qualitatively different cases;

\item we find analytically isotropic exponential solutions: we substitute $\dot H = \dot h \equiv 0$ as well as $h = H$ into (\ref{H_gen})--(\ref{con2_gen}); the system simplifies into a single equation,
we solve it and find not only roots but also the ranges of $(\alpha, \beta)$ where they exist;

\item we find analytically anisotropic exponential solutions: we substitute \linebreak \mbox{$\dot H = \dot h \equiv 0$} into (\ref{H_gen})--(\ref{con2_gen}); the system could be brought down to two equations: bi-six order polynomial
in $h$ with powers of $\alpha$ and $\beta$ as coefficients and
$H = H(h, \alpha, \beta)$. Both of them are usually higher-order with respect to their arguments so retrieving the solutions in radicand is impossible. But if we consider the discriminant of the former of them, the resulting
equation gives us critical values for $(\alpha, \beta)$ which separate areas with different number of roots;

\item first three steps provides us with a set of critical values for $(\alpha, \beta)$ which separate domains with different dynamics;

 \item we solve (\ref{H_gen})--(\ref{h_gen}) with respect to $\dot H$ and $\dot h$;

 \item we substitute obtained $H_i$ into $\dot H$ and $\dot h$ and obtain the latter as a single-variable functions: $\dot H(h)$ and $\dot h(h)$;

 \item the obtained $\dot H(h)$ and $\dot h(h)$ expressions and graphs are analyzed for all possible domains in $(\alpha, \beta)$ space to obtain all possible regimes;

 \item obtained exponential regimes are compared with exact isotropic and anisotropic solutions (see~\cite{CPT3}) to find the nature of the exponential regimes in question;

 \item power-law regimes are analyzed in terms of Kasner exponents ($p_i = - H_i^2/\dot H_i$) to verify that low-energy power-law regimes are standard Kasner regimes with
 \mbox{$\sum p_i = \sum p_i^2 = 1$} or ``generalized Taub'' regimes (see below) while high-energy
 power-law regimes are Lovelock Kasner regimes with \linebreak \mbox{$\sum p_i = (2n-1) = 5$}.

\end{itemize}

The above scheme allows us to completely describe all existing regimes for a given set of the parameters $(\alpha, \beta)$. In the previous paper~\cite{my18b} we followed the scheme in great detail for $D=3$ case, with full description of all steps, and in the current papers we just briefly mentioned the details and mainly focus on the results.

Before proceeding with the particular cases, it is useful to introduce the notations we are going to use through the paper. We denote Kasner regime as $K_i$ where $i$ is the
total expansion rate in terms of the Kasner exponents $\sum p_i = (2n-1)$ where $n$ is the corresponding order of the Lovelock contribution (see, e.g.,~\cite{prd09}).
So that for Einstein-Hilbert contribution $n=1$ and $\sum p_i = 1$ (see~\cite{kasner}) and the corresponding regime is $K_1$, which is usual low-energy regime in vacuum
EGB case (see~\cite{my16a, my18a}) and we expect it to remain here. For Gauss-Bonnet $n=2$ and so $\sum p_i = 3$ and the regime $K_3$ is typical high-energy regime for EGB case (again, see~\cite{my16a, my16b, my17a, my18a}). Finally, for cubic Lovelock $n=3$ and so $\sum p_i = 5$ and the regime $K_5$ is typical high-energy regime for
this case in low $D$ case~\cite{my18b}.

Another power-law regime is what is called ``generalized Taub'' (see~\cite{Taub} for the original solution).
We mistakenly taken it for $K_3$ in~\cite{my16a}, but then in~\cite{my18a}
corrected ourselves and explained the details (they both have $\sum p_i = 3$ which causes misinterpreting). It is a situation when for one of the subspaces the Kasner exponent
$p$ is equal to zero and for another --
to unity. So we denote
$P_{1, 0}$ the case with $p_H = 1, p_h = 0$ and $P_{0, 1}$ the case with $p_H = 0, p_h = 1$.

The exponential solutions are denoted as $E$ with subindex indicating its details -- $E_{iso}$ is isotropic exponential solution and $E_{3+D}$ is anisotropic -- with different Hubble
parameters corresponding to three- and extra-dimensional subspaces. But in practice, in each particular case there are several different anisotropic exponential solutions, so that instead
of using $E_{3+D}$ we use $E_i$ where $i$ counts the number of the exponential solution ($E_1$, $E_2$ etc). In case if there are several isotropic exponential solutions, we count them with upper index:
$E_{iso}^1$, $E_{iso}^2$ etc.

And last but not least regime is what we call ``nonstandard singularity'' and denote it is as $nS$.
It is the situation which arise in Lovelock gravity due to its nonlinear nature. Since the equations
(\ref{H_gen})--(\ref{h_gen}) are nonlinear with respect to the highest derivative ($\dot H$ and $\dot h$ in our case), when we solve them, the resulting expressions are ratios with
polynomials in both numerator and denominator. So there exist a situation when the denominator is equal to zero for finite values of $H$ and\/or $h$. This situation is singular, as the
curvature invariants diverge, but it happening for finite values of $H$ and\/or $h$. Tipler~\cite{Tipler} call this kind of singularity as ``weak'' while Kitaura and
Wheeler~\cite{KW1, KW2} -- as ``type II''. Our previous research demonstrate that this kind of singularity is widely spread in EGB cosmology -- in particular, in totally anisotropic
(Bianchi-I-type) $(4+1)$-dimensional vacuum cosmological model it is the only future asymptote~\cite{prd10}.

\section{$D=5$ case}

In this case the equations of motion (\ref{H_gen})--(\ref{con2_gen}) take form ($H$-equation, $h$-equation, and constraint correspondingly)

\begin{equation}
\begin{array}{l}
4\dot H + 6H^2 + 10\dot h + 30h^2 + 20Hh + 8\alpha \( 2\dot H (5Hh + 10h^2) + 5\dot h (H^2 + 6h^2 + 8Hh) + \right.\\ \left. + 10H^3h + 55H^2h^2  + 100Hh^3 + 45h^4 \) + 144\beta \( 2 (\dot H + H^2)(10Hh^3 + 5h^4) + \right.\\ \left.
+ 5(\dot h+ h^2)(6H^2h^2 + 8Hh^3 + h^4)   + 25H^2h^4 + 10Hh^5   \)
 = 0,
\end{array} \label{D5_H}
\end{equation}

\begin{equation}
\begin{array}{l}
6\dot H + 12H^2 + 8\dot h + 20h^2 + 24Hh + 8\alpha \( 3\dot H (H^2 + 8Hh + 6h^2) + 4\dot h (3H^2 + 9Hh + 3h^2)
+ \right. \\ \left.
+ 3H^4 + 36H^3h  + 84H^2h^2 + 72Hh^3 + 15h^4 \) + 144\beta \( 3(\dot H + H^2) (6H^2h^2 + 8Hh^3 + h^4)
+ \right. \\ \left. + 4(\dot h + h^2)(3H^3h + 9H^2h^2 + 3Hh^3) +20H^3h^3 +15H^2h^4 \)
 = 0,
\end{array} \label{D5_h}
\end{equation}

\begin{equation}
\begin{array}{l}
6 H^2 + 30Hh + 20h^2 + 24\alpha (5H^3h + 30H^2h^2 + 30Hh^3 + 5h^4 ) + \\  + 720\beta ( 10H^3h^3 + 15H^2h^4 + 3Hh^5) = 0.
\end{array} \label{D5_con}
\end{equation}

Following the described above procedure, first we find the critical values for $\mu$ for $H(h)$. We find the discriminant of (\ref{D5_con}) with respect to $H$ and then its discriminant with respect to $h$ -- it is 16th-order
polynomial with respect to $\mu = \beta/\alpha^2$ and it has the following roots: double root

$$
\mu_1 = - \dac{\sqrt[3]{2150 + 210\sqrt{105}}}{210} + \dac{2}{21\sqrt[3]{2150 + 210\sqrt{105}}} + \dac{11}{42} \approx 0.1903,
$$

\noindent quadruple root $\mu_2 = 5/8 = 0.625$ and
sextuple root $5/14$; the last one does not affect the dynamics.

The next step is the abundance of the isotropic exponential solutions; we substitute $\dot H = \dot h \equiv 0$ and $h \equiv H$ into (\ref{D5_H})--(\ref{D5_con}) to obtain single equation which governs the isotropic
solutions:

$$
56 H^2 + 1680\alpha H^4 + 20160\beta H^6 = 0.
$$

\noindent The analysis of its nontrivial solutions

$$
H^2 = \dac{-5\alpha \pm \sqrt{25\alpha^2 - 50\beta} }{120\beta}
$$

\noindent suggests that we have one isotropic exponential solution iff $\beta < 0$ (for any sign of $\alpha$) and two
isotropic exponential solutions for $\alpha < 0$, $\beta > 0$, $\mu \leqslant 5/8$.

The final step before the regime analysis is the abundance of the anisotropic exponential solutions. We substitute $\dot H = \dot h \equiv 0$ into (\ref{D5_H})--(\ref{D5_con}) and solve the resulting system with
respect to $H$ and $h$. The resulting equation on $h$ is bi-9-power and its discriminant is 24th-order polynomial on $\mu$ with the following roots: two roots $\mu_3 \approx 0.0517, \mu_4 \approx 0.1328$ belong to a
certain six-order polynomial and cannot be expressed through elementary functions; double pair of roots $\mu_{5, 6} = 55/18 \pm 10\sqrt{7}/9 \approx 0.1158, 5.9953$; single roots $\mu_7 = 1681/8242 \approx 0.1995$ and
$\mu = 5/8 \equiv \mu_2$; plus there is a triple set of imaginary roots from a fourth-order polynomial. The further analysis suggests that for $\alpha < 0$, $\beta < 0$ there is one anisotropic exponential solution,
for $\alpha < 0$, $\beta > 0$ there are two anisotropic exponential solutions if $\mu > 5/8$. For $\alpha > 0$, $\beta < 0$ there are also two anisotropic exponential solutions while for $\alpha > 0$, $\beta > 0$ the
situation is more complicated, similar to the previous cases. So for $\mu < \mu_3$ there are three, for $\mu = \mu_3$ four and for $\mu_4 > \mu > \mu_3$ five. With further growth of $\mu$ we have decrease of the number of solutions: four for $\mu = \mu_4$, three for $\mu_7 > \mu > \mu_4$, two for $\mu = \mu_7$ and only one for $\mu > \mu_7$. One can see that, similar to the previous cases, there is a fine structure of the anisotropic
exponential solutions at $\alpha > 0$, $\beta > 0$ and we describe it separately.

\begin{figure}
\centering
\includegraphics[width=0.7\textwidth, angle=0]{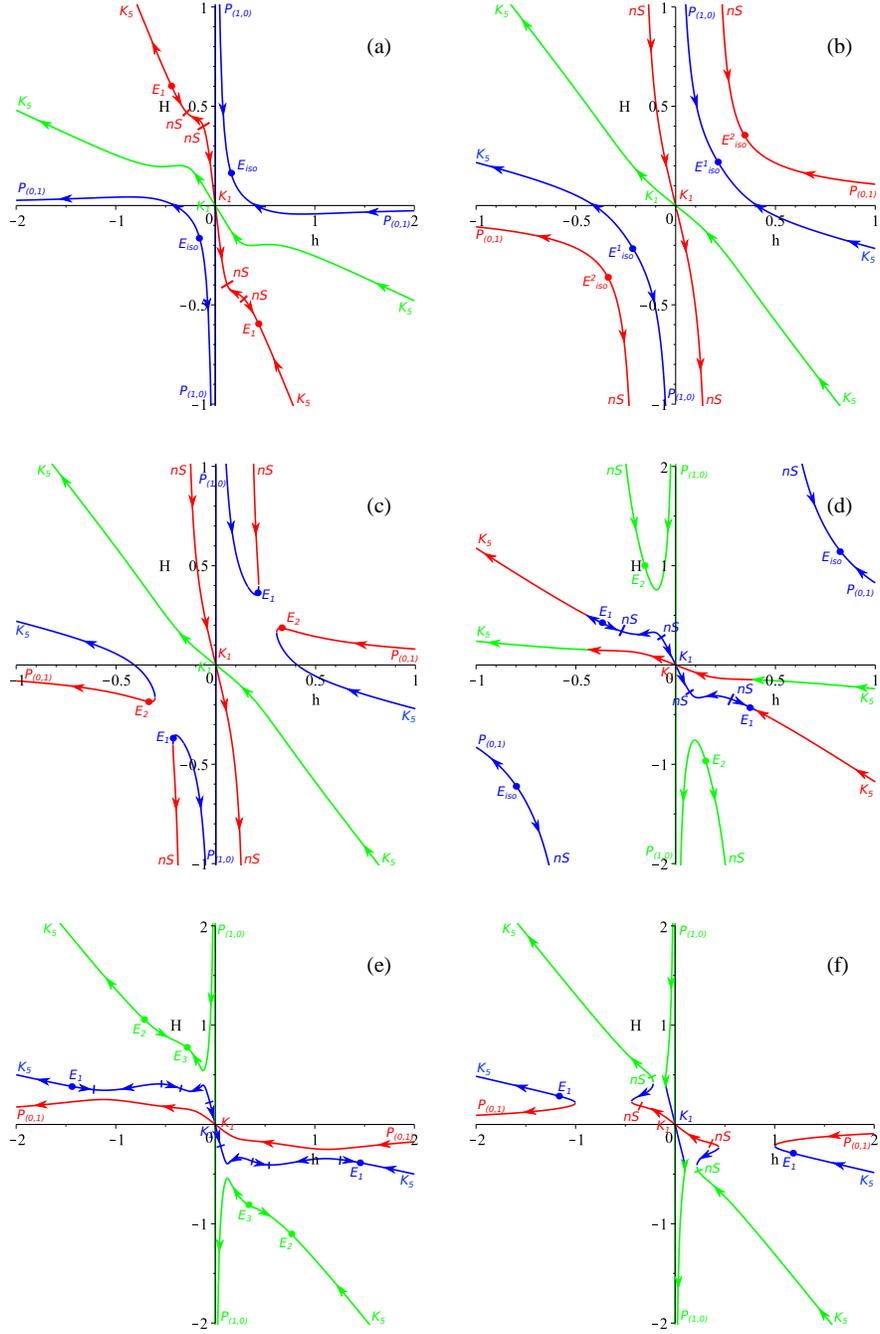}
\caption{Final compilations of all possible regimes in $D=5$ vacuum cubic Lovelock case, on $H(h)$ evolution curves; different colors correspond to three different branches $H_1$, $H_2$ and $H_3$;
panels layout is as follows: $\alpha < 0$, $\beta < 0$ on (a) panel, $\alpha < 0$, $\beta > 0$, $\mu < 5/8$ on (b) panel, $\alpha < 0$, $\beta > 0$, $\mu > 5/8$ on (c) panel, $\alpha > 0$, $\beta < 0$ on
(d) panel, $\alpha > 0$, $\beta > 0$, $\mu < \mu_3$ on (e) panel, and $\alpha > 0$, $\beta > 0$, $\mu > \mu_7$ on (f) panel
(see the text for more details).}\label{D5_1}
\end{figure}

With all preliminaries done, it is time to describe all individual regimes. We solve (\ref{D5_con}) with respect to $H$ and plot the resulting $H(h)$ curves in Fig.~\ref{D5_1}.
There red curve corresponds to $H_1$, blue to $H_2$ and green
to $H_3$. The panels layout is as follows: $\alpha < 0$, $\beta < 0$ on (a) panel, $\alpha < 0$, $\beta > 0$, $\mu < 5/8$ on (b) panel, $\alpha < 0$, $\beta > 0$, $\mu > 5/8$ on (c) panel, $\alpha > 0$, $\beta < 0$ on
(d) panel, $\alpha > 0$, $\beta > 0$, $\mu < \mu_3$ on (e) panel, and $\alpha > 0$, $\beta > 0$, $\mu > \mu_7$ on (f) panel.

We analyze the individual $\dot H(h)$, $\dot h(h)$, $p_H$, $p_h$ curves and plot the derived regimes directly on $H(h)$ evolution curves (see Fig.~\ref{D5_1}). The resulting regimes for $\alpha < 0$, $\beta < 0$,
presented in Fig.~\ref{D5_1}(a), are: $P_{1, 0} \toto E_{iso}$ and $P_{0, 1} \toto E_{iso}$ on blue hyperbola-like $H_2$ curve, $K_5 \toto K_1$ on green $H_3$ curve and $K_5 \toto E_1$, $nS \toto E_1$, $nS \toto nS$
and $K_1 \toto nS$ on $H_1$. One can see that neither of the regimes have compactification feature -- $K_1$ from $H_3$ has $H < 0$, $K_1$ from $H_1$ has $nS$ as past attractor, $E_1$ with $H > 0$ is unstable.

The next case to consider is $\alpha < 0$, $\beta > 0$, $\mu < 5/8$, presented in Fig.~\ref{D5_1}(b). The regimes there are: $P_{0, 1} \toto E_{iso}^2$ and $nS \toto E_{iso}^2$ on hyperbola-like part of $H_1$,
$nS \toto K_1$ on the remaining part of $H_1$, $P_{1, 0} \toto E_{iso}^1$ and $K_5 \toto E_{iso}^1$ on hyperbola-like $H_2$, and $K_5 \toto K_1$ on $H_3$. For the same reasons as in the previous case, $K_1$ cannot be called
realistic compactifications, and since there are no other candidates, there are no compactifications in this case either.

Previous case naturally followed by $\alpha < 0$, $\beta > 0$, $\mu > 5/8$, presented in Fig.~\ref{D5_1}(c). On the boundary value, $\mu = 5/8$, the isotropic exponential solutions from Fig.~\ref{D5_1}(b) coincide so that
we have one pair instead of two, but the other regimes are the same so we skipped $\mu = 5/8$ from separate consideration. The $\mu > 5/8$ case has the regimes: $K_5 \toto K_1$ on $H_3$ and $nS \toto K_1$ like in previous case,
$E_1 \toto nS$ and $E_1 \toto P_{1, 0}$ on one banana-like curve and $E_2 \toto P_{0, 1}$ and $E_2 \toto K_5$ on another banana-like curve. For the same reasons as in the previous cases, $K_1$ cannot give us realistic compactification,
and $E_1$ and $E_2$ are located either in first, so that having $H > 0$, $h > 0$, or in third ($H < 0$, $h < 0$) quadrants, so that also cannot serve as realistic compactification
either. For $\alpha < 0$, similar to the previous
$D$ cases (see~\cite{my18b}), there are no realistic compactification regimes.

We proceed with $\alpha > 0$, and the first case to consider is $\beta < 0$, presented in  Fig.~\ref{D5_1}(d). There on a hyperbola-like part of $H_2$ branch we have $P_{0, 1} \toto E_{iso}$ and $nS \toto E_{iso}$ and
$P_{1, 0} \toto E_2$ and $nS \toto E_2$ on edge-shaped part of $H_3$ branch. On the latter, $P_{1, 0} \to E_2$ is viable compactification regime -- $E_2$ has $H > 0$ and $h < 0$ and is stable past asymptote. The remaining
regimes include $K_5 \toto K_1$ which possesses features of the previous cases and so has not compactifications, and the following combination of regimes along $H_1-H_2$ physical branch: $K_5 \to E_1 \ot nS \to nS \ot K_1$.
The regimes are given according to fourth quadrant, in the second quadrant they are time-reversed. For the similar reasons as in the previous cases, neither of the regimes along $H_1-H_2$ physical branch have realistic
compactification.

\begin{figure}
\centering
\includegraphics[width=0.96\textwidth, angle=0]{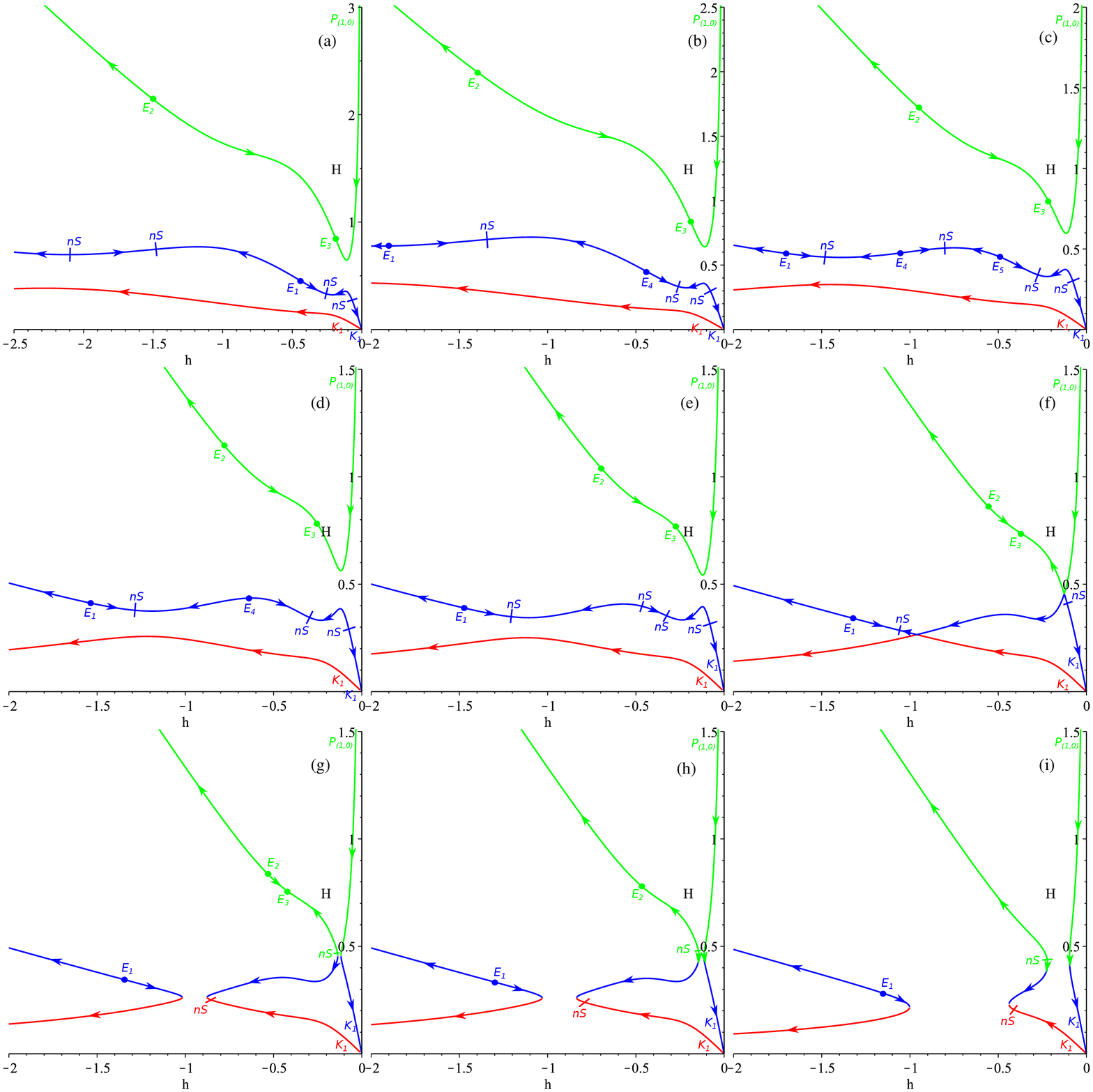}
\caption{The fine structure of the solutions in the $\alpha > 0$, $\beta > 0$ $D=5$ vacuum case:
$\mu < \mu_3$ on (a) panel, $\mu = \mu_3$ on (b) panel, $\mu_4 > \mu > \mu_3$ on (c) panel,
$\mu = \mu_4$ on (d) panel, $\mu_1 > \mu > \mu_4$ on (e) panel, $\mu = \mu_1$ on (f) panel,
$\mu_7 > \mu > \mu_1$ on (g) panel, $\mu = \mu_7$ on (h) panel and $\mu > \mu_7$ on (i) panel.
Different colors correspond to the different branches (red -- to $H_1$, blue -- to $H_2$ and green -- to $H_3$, in accordance with the designation in Fig.~\ref{D5_1})
(see the text for more details).}\label{D5_2}
\end{figure}

The final case to consider is $\alpha > 0$, $\beta >0$. The limiting cases -- $\mu < \mu_3$ and $\mu > \mu_7$ are presented in Fig.~\ref{D5_1} -- Fig.~\ref{D5_1}(e) for $\mu < \mu_3$ and Fig.~\ref{D5_1}(f) for $\mu > \mu_7$;
the fine structure between them is presented in Fig.~\ref{D5_2}: $\mu < \mu_3$ (the same as in Fig.~\ref{D5_1}(e) but detailed range) on (a) panel, $\mu = \mu_3$ on (b) panel, $\mu_4 > \mu > \mu_3$ on (c) panel,
$\mu = \mu_4$ on (d) panel, $\mu_1 > \mu > \mu_4$ on (e) panel, $\mu = \mu_1$ on (f) panel, $\mu_7 > \mu > \mu_1$ on (g) panel, $\mu = \mu_7$ on (h) panel and $\mu > \mu_7$ (the same as in Fig.~\ref{D5_1}(f) but detailed
range) on (i) panel.

Let us now describe the regimes which appear in these cases. At $\mu < \mu_3$ (see Figs.~\ref{D5_1}(e) and~\ref{D5_2}(a)) we have $K_1 \to K_5$ along the $H_1$ branch (we focus on the second quadrant; the regimes in the fourth quadrant are time-reversal of the described), \linebreak  $K_1 \ot nS \to nS \ot E_1 \to nS \ot nS \to K_5$ along $H_2$ and $P_{1, 0} \to E_3 \ot E_2 \to K_5$ along $H_3$. Of these regimes only $P_{1, 0} \to E_3$ from the $H_3$ branch
has realistic compactification. For the next several cases the regimes along $H_1$ and $H_3$ do not change (and so the realistic compactification $P_{1, 0} \to E_3$ from the $H_3$ branch is presented in them as well), the
difference is only within $H_2$ branch and we describe only these changes. This way, for $\mu = \mu_3$ (Fig.~\ref{D5_2}(b)) along $H_2$ we have $K_1 \ot nS \to nS \ot E_4 \to nS \ot E_1 \to K_5$ (the outmost $nS$ turned into
anisotropic exponential solution), but no new viable compactifications appear; for $\mu_4 > \mu > \mu_3$ (Fig.~\ref{D5_2}(c)) we have $K_1 \ot nS \to nS \ot E_5 \to nS \ot E_4 \to nS \ot E_1 \to K_5$; for for $\mu = \mu_4$ (Fig.~\ref{D5_2}(d)) we have $K_1 \ot nS \to nS  \ot E_4 \to nS \ot E_1 \to K_5$; for $\mu_1 > \mu > \mu_4$ (Fig.~\ref{D5_2}(e)) we have $K_1 \ot nS \to nS \ot   nS    \to nS \ot E_2 \to K_5$. One can see that neither of
the regimes which appear on $H_2$ have realistic compactification. The next case is $\mu = \mu_1$, presented in  Fig.~\ref{D5_2}(f) and this is the situation when the physical branches ``touch'' each other and ``reconnecting'',
making different ``routes'', so that many of the existing regimes terminating and other regimes are formed instead. New regimes could be found on Fig.~\ref{D5_2}(g), where we plot $\mu_7 > \mu > \mu_1$ case. One can note that
the abundance of the exponential solutions and nonstandard singularities is exactly the same as for $\mu_1 \geqslant \mu > \mu_4$ (Figs.~\ref{D5_2}(e, f)) but the regimes are completely different, since $H(h)$ forming completely
different physical branches. So the new regimes for $\mu_7 > \mu > \mu_1$  (Fig.~\ref{D5_2}(g))
are: $K_5 \ot E_1 \to K_5$ on the outmost branch, $K_5 \ot E_2 \to E_3 \ot nS \to nS \ot K_1$ on the middle branch and $P_{1, 0} \to K_1$ on the innermost branch. One can see that
only the
latter gives us realistic compactification -- in all other cases $K_5$ is future asymptote (or past asymptote but for the regimes in the fourth quadrant, which have $H < 0$). Further increase of $\mu$ changing the regimes
along middle branch, leaving the outmost and innermost unchanged and so keeping $P_{1, 0} \to K_1$ as a viable compactification. The change of the regimes along the middle branch is as follows:
$K_5 \ot E_2  \ot nS \to nS \ot K_1$ for $\mu = \mu_7$ (see Fig.~\ref{D5_2}(h)) and $K_5 \ot nS \to nS \ot K_1$ for $\mu > \mu_7$ (see Fig.~\ref{D5_2}(i)); one can see that there are no realistic compactification regimes among those
within $H_2$.

To conclude, the amount of the viable compactification regimes exactly the same as in $D=4$ case (see~\cite{my18b}) -- in both cases we have $P_{1, 0} \to E_{3+D}$ for $\alpha > 0$, $\beta < 0$ and $P_{1, 0} \to E_{3+D} / K_1$ for
$\alpha > 0$, $\beta > 0$, $\mu < \mu_1$ or $\mu > \mu_1$ where $\mu_1$ is separating value for $H(h)$. And again, for entire $\alpha > 0$ we have realistic compactification regimes.

\section{General $D \geqslant 6$ case}

In this case we use the core (\ref{H_gen})--(\ref{con2_gen}) equations -- for all $D \geqslant 6$ the structure of the equations of motion is unchanged. Following the procedure, we find the discriminant of (\ref{con2_gen})
with respect to $H$ and then its discriminant with respect to $h$ -- it is 16th-order
polynomial with respect to $\mu = \beta/\alpha^2$ and it has the following roots:

$$
\mu_1 = \dac{D(D+1)}{4(D-1)(D-2)}
$$

\noindent and $\mu_2 < \mu_3$ -- the solutions of certain six-order polynomial with the coefficients made up to $D^{16}$.
The expressions for $\mu_{2,3}$ cannot be obtained in terms of elementary functions in the general case but for each $D$ they could be derived, at least numerically. Similar to the previous cases, $\mu_1$ separates two
different $H(h)$ regimes in $\alpha < 0$, $\beta > 0$ while $\mu_{2, 3}$ -- for $\alpha > 0$, $\beta > 0$, Unlike previous $D=3, 4$ cases (see~\cite{my18b}), where there is only one separation $\mu$ in the $\alpha > 0$, $\beta > 0$ case, now
we have two.

Substituting $\dot H = \dot h \equiv 0$ and $h \equiv H$ into (\ref{H_gen})--(\ref{con2_gen}) gives us single equation which governs isotropic exponential solutions:

$$
H^2(D+2)(D+3) + \alpha H^4 D(D+1)(D+2)(D+3) + \beta H^6 (D-2)(D-1)D(D+1)(D+2)(D+3) = 0.
$$

\noindent Analysis of its nontrivial solution

$$
H^2 = - \dac{ \alpha D(D+1) \pm \sqrt{ \alpha D^2 (D+1)^2 - \beta D (D-2)(D-1)(D+1) } }{\beta D(D-2)(D-1)(D+1)}
$$

\noindent suggests that there is one isotropic exponential solution for $\beta < 0$
(for both signs of $\alpha$) and two for $\alpha < 0$, $\beta > 0$, $\mu < \mu_1$. Let us note that this is the same scheme we had for all previous cases, making it true for all $D$.

The situation with anisotropic exponential solutions is more complicated. Following usual procedure, we obtain equation for $h$ as bi-nine-power polynomial with the discriminant being 24th-order polynomial in $\mu$ with
coefficients made up to $D^{130}$. Interesting enough, in the general case there are only two roots of this discriminant which affect the structure of the anisotropic exponential solutions: $\mu_1$ -- the same as
in $H(h)$ and isotropic exponential solutions -- and

$$
\mu_4 = - \dac{D^4 + 30G^3 + 189 D^2 - 540 D + 324}{D^4 - 6D^3 - 25D^2 + 102D - 72}.
$$

\noindent So that we can say that in the general $D \geqslant 6$ case there is no ``fine structure'' of the anisotropic exponential solutions -- at least not in the sense we have it for $D=3, 4, 5$. One can also note that
$\mu_{2, 3, 4}$ have different locations at different $D$: $\mu_3 > \mu_4 > \mu_2 > 0$ for $D=6$, $\mu_4 > \mu_3 > \mu_2 > 0$ for $D=7$ and $\mu_4 < 0$ for  $D \geqslant 8$. So that these three cases have slightly different
dynamics and we decided to describe them separately. We fully describe $D=6$ and then describe the changes and new regimes which come from $D=7$ and $D \geqslant 8$.

\subsection{$D=6$ case}

First let us write down the values for $\mu_i$ in this case. So $\mu_1 = 21/40 = 0.525$, $\mu_2 \approx 0.2151$, $\mu_3 \approx 0.3382$ and $\mu_4 = 0.3$.

\begin{figure}
\centering
\includegraphics[width=0.7\textwidth, angle=0]{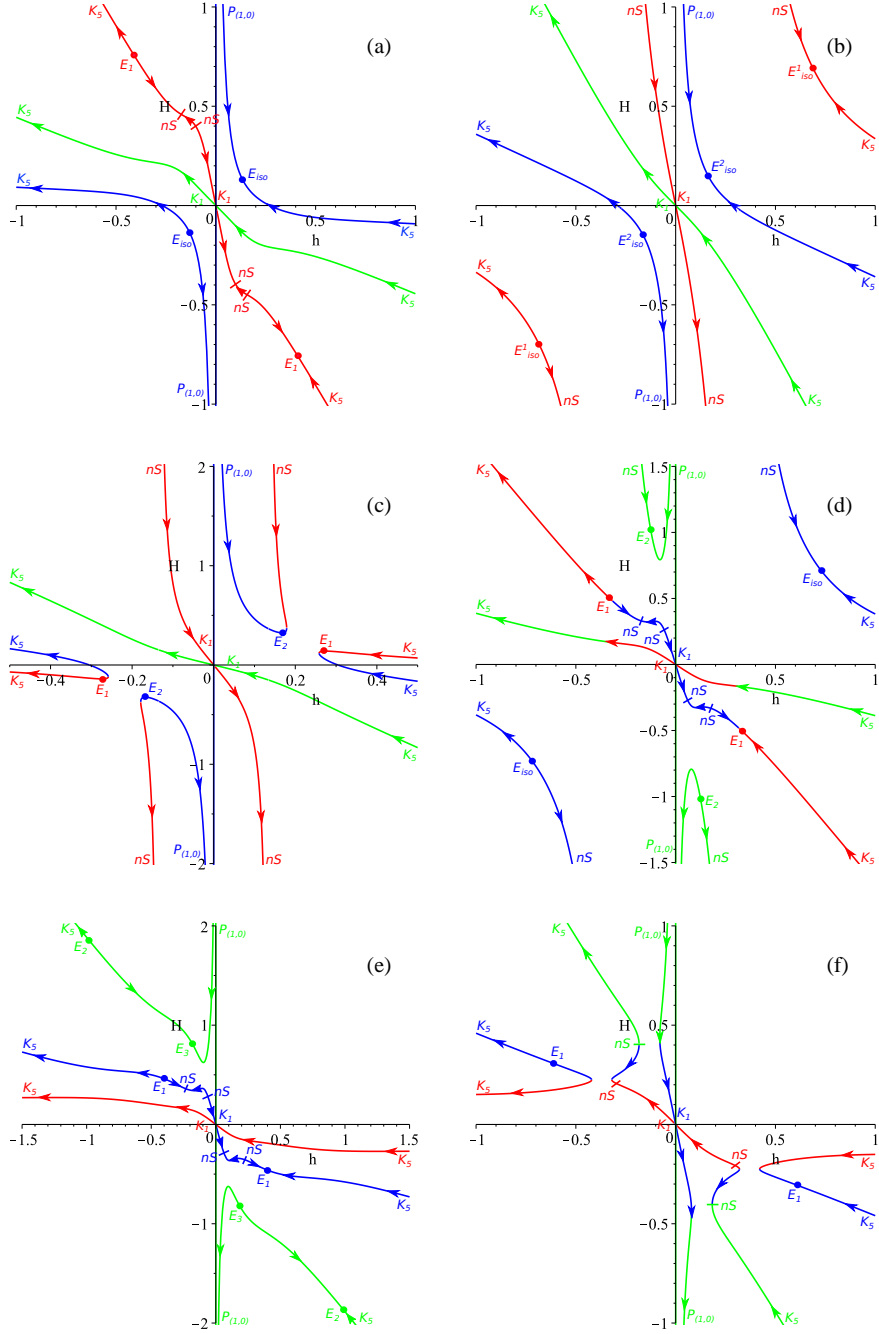}
\caption{Final compilations of all possible regimes in $D=6$ vacuum cubic Lovelock case, on $H(h)$ evolution curves; different colors correspond to three different branches $H_1$, $H_2$ and $H_3$;
panels layout is as follows: $\alpha < 0$, $\beta < 0$ on (a) panel, $\alpha < 0$, $\beta > 0$, $\mu < \mu_1$ on (b) panel, $\alpha < 0$, $\beta > 0$, $\mu > \mu_1$ on (c) panel, $\alpha > 0$, $\beta < 0$ on
(d) panel, $\alpha > 0$, $\beta > 0$, $\mu < \mu_2$ on (e) panel, and $\alpha > 0$, $\beta > 0$, $\mu > \mu_3$ on (f) panel
(see the text for more details).}\label{D6_1}
\end{figure}

We perform the same procedure as in the previous cases and obtain the resulting regimes on the $H(h)$ curves; we present them in Fig.~\ref{D6_1}. As always, different colors correspond to different branches:
$H_1$ is red, $H_2$ is blue and $H_3$ is green. The panels layout is the following: $\alpha < 0$, $\beta < 0$ presented on (a) panel, $\alpha < 0$, $\beta > 0$, $\mu < \mu_1$ on (b) panel, $\alpha < 0$, $\beta > 0$, $\mu > \mu_2$ on (c) panel, $\alpha > 0$, $\beta < 0$ on
(d) panel, $\alpha > 0$, $\beta > 0$, $\mu < \mu_2$ on (e) panel, and $\alpha > 0$, $\beta > 0$, $\mu > \mu_3$ on (f) panel. Let us now describe the regimes.

The first case to consider, $\alpha < 0$, $\beta < 0$, is presented in Fig.~\ref{D6_1}(a). There one can see $K_5 \toto E_{iso}$ and $P_{1, 0} \toto E_{iso}$ on hyperbola-like $H_2$ branch, $K_5 \toto K_1$ on $H_3$ and
a number of regimes along $H_1$ (according to the second quadrant; in fourth quadrant they are time-reversed): $K_5 \ot E_1 \to nS \ot nS \to K_1$. One can see that neither of the listed regimes have viable compactification:
$E_1$ is either unstable or has $H < 0$, as for $K_1$, it is either past attractor or has $H < 0$, or has $nS$ as a past attractor.

The next case is $\alpha < 0$, $\beta > 0$, $\mu < \mu_1$ and it is presented in Fig.~\ref{D6_1}(b). There we have $K_5 \toto E_{iso}^1$ and $nS \toto E_{iso}^2$ on hyperbola-like part of $H_1$ branch and $nS \toto K_1$ on the
remaining part, $K_5 \toto E_{iso}^2$ and $P_{1, 0} \toto E_{iso}^2$ on $H_2$ and $K_5 \toto K_1$ on $H_3$. Similar to the previous case (and for the similar reasons) there are no realistic compactification schemes in this case as well.

Similarly to previously considered $D$ cases, at $\mu = \mu_1$ the isotropic exponential solutions $E_{iso}^1$ and $E_{iso}^2$ ``touch'' each other and coincide; the regimes remain the same so we skip this case from consideration.

With further increase of $\mu$ we have the next case -- $\alpha < 0$, $\beta > 0$, $\mu > \mu_1$, presented in Fig.~\ref{D6_1}(c). Again, similar to the previous $D$ cases, isotropic exponential solutions switched into anisotropic ones; hyperbola-like branches, which ``touch'' each other at $\mu = \mu_1$, ``detouch'' and form new physical ``banana-shaped'' branches with the exponential solutions are located on them -- one at a branch.
So on one of them we have $K_5 \toto E_1$ with two different $K_5$, on another $nS \toto E_2$ and $P_{1, 0} \toto E_2$; there are also $nS \toto K_1$ and $K_1 \toto K_5$ on $H_3$ -- the sane as in previous cases. One can see that $E_1$ and $E_2$ in the third quadrant are unstable while those in the first are stable but have $H > 0$ and $h > 0$ so they cannot describe realistic compactification\footnote{At large $\mu$, $E_1$ could have $H < 0$ and move into the fourth quadrant, but still could not be called viable compactification.}. The comments about $K_1$ from the previous cases are true here as well, so there are no viable compactification in this case. To conclude,
similarly to the previous $D$ cases, $\alpha < 0$ domain does not have realistic compactification regimes.

Now let us turn to $\alpha > 0$. The first of these cases is $\beta < 0$, presented in Fig.~\ref{D6_1}(d). The regimes for this case are: $K_5 \toto E_{iso}$ and $nS \toto E_{iso}$ on hyperbola-like $H_2$ part, $P_{1, 0} \toto E_2$
and $nS \toto E_2$ on the edge-shaped $H_3$ part; among them $P_{1, 0} \to E_2$ is a realistic compactification regime. Other regimes include $K_5 \to K_1$ on one and $K_5 \ot E_1 \to nS \ot nS \to K_1$ on another physical
branch, listed according to the second quadrant. Let us note that the listed regimes along the second branch are exactly the same as in the described above $\alpha < 0$, $\beta < 0$ case (see Fig.~\ref{D6_1}(a)). So that in
this case we have one viable compactification regime -- $P_{1, 0} \to E_2$.

\begin{figure}
\centering
\includegraphics[width=0.75\textwidth, angle=0]{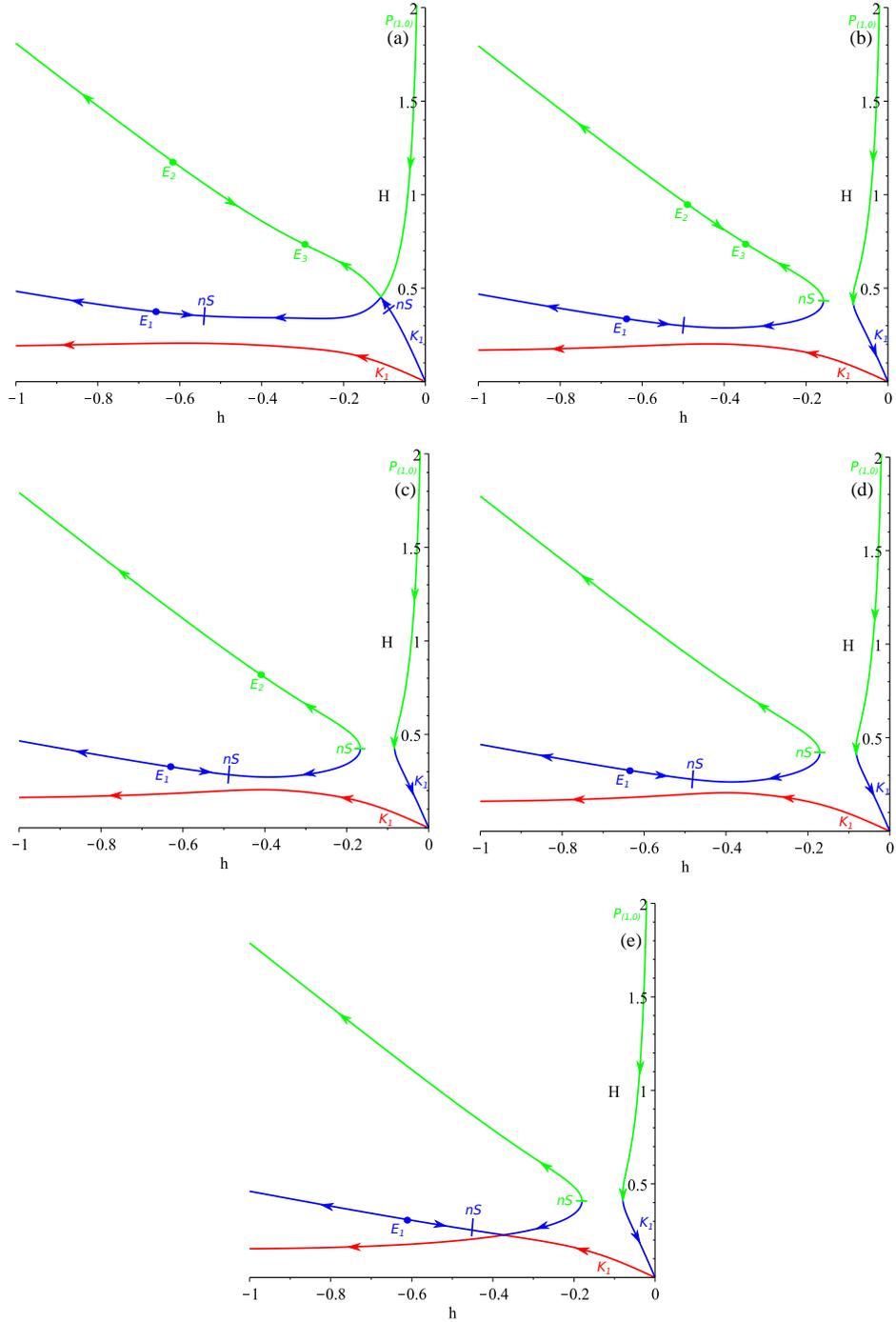}
\caption{The fine structure of the solutions in the $\alpha > 0$, $\beta > 0$ $D=6$ vacuum case:
$\mu = \mu_2$ on (a) panel, $\mu_4 > \mu > \mu_2$ on (b) panel, $\mu = \mu_4$ on (c) panel, $\mu_3 > \mu > \mu_4$ on (d) panel
and $\mu = \mu_3$ on (e) panel. Different colors correspond to different branches: $H_1$ is red, $H_2$ is blue and $H_3$ is green, in accordance with Fig.~\ref{D6_1}
(see the text for more details).}\label{D6_2}
\end{figure}

What remains is the description of the cases for $\alpha > 0$, $\beta > 0$. The first of the $\alpha > 0$, $\beta > 0$ cases, $\mu < \mu_2$, is presented in Fig.~\ref{D6_1}(e), the following cases -- in Fig.~\ref{D6_2} and
the last case, $\mu > \mu_3$, in Fig.~\ref{D6_1}(f). In this regard, we called the subcases in Fig.~\ref{D6_2} as a ``fine-structure'' of the $\alpha > 0$, $\beta > 0$ case, but it is not in the same sense as previous $D$
cases. In the previous $D$ cases we have fine-structure with respect to the rapid change in the number of exponential solutions -- within a small $\mu$ region, the exponential solutions appear, disappear, merge, turn in $nS$ -- they
have rich and rapid-changing dynamics. In this case we does not have all of this, but we have change of the physical branches. So the manner is the same and so we keep the name ``fine-structure''. The subcases within it are
presented in Fig.~\ref{D6_2} and the panel layout is as follows: $\mu = \mu_2$ is on (a) panel, $\mu_4 > \mu > \mu_2$ is on (b) panel, $\mu = \mu_4$ is on (c) panel, $\mu_3 > \mu > \mu_4$ is on (d) panel and $\mu = \mu_3$
is on (e) panel.

Let us start with $\mu < \mu_2$ case, presented in Fig.~\ref{D6_1}(e). The regimes for this case are (we list them according to the second quadrant): $K_1 \to K_5$ on $H_1$, \mbox{$K_5 \ot E_1 \to nS \ot nS \to K_1$} on $H_2$ and
$K_5 \ot E_2 \to E_3 \ot P_{1, 0}$. Of all the regimes, only $P_{1, 0} \to E_3$ have realistic compactification. The next case is $\mu = \mu_2$, presented in Fig.~\ref{D6_2}(a). Here one can immediately see the difference
between $D=4, 5$ and the general $D \geqslant 6$ cases -- one can see that there are two touching points for the branches and in $D=4, 5$ they touch and decouple at the same $\mu$ (see e.g. Fig.~\ref{D5_2}(f)). But in the
general $D \geqslant 6$ they touch and decouple at two different $\mu$ -- one at a time -- and this is the reason for ``fine-structured'' consideration of the $\alpha > 0$, $\beta > 0$ case. The regimes in this case are the
same as for $\mu < \mu_2$. Once $\mu$ is increased to $\mu_4 > \mu > \mu_2$, the regimes changes and are presented in Fig.~\ref{D6_2}(b). The regimes are: $P_{1, 0} \to K_1$, $K_1 \to K_5$ and
$K_5 \ot E_2 \to E_3 \ot nS \to nS \ot E_1 \to K_5$. Of them, only $P_{1, 0} \to K_1$ has realistic compactification. The next case is $\mu=\mu_4$, presented in Fig.~\ref{D6_2}(c), and at this point the number of the
exponential solution changes. So that
instead of a pair of anisotropic exponential solutions $E_2$ and $E_3$ on $H_3$, we have singe meta-stable ($\sum H = 0$ -- constant-volume solution, see~\cite{CST2}) $E_2$. The $P_{1, 0} \to K_1$ and $K_1 \to K_5$ regimes remain unchanged
and the regimes list with anisotropic exponential solution is: $K_5 \ot E_2 \ot nS \to nS \ot E_1 \to K_5$. The same as in previous case, only $P_{1, 0} \to K_1$ has realistic compactification. With further increase of $\mu$,
for $\mu_3 > \mu > \mu_4$, presented in Fig.~\ref{D6_2}(d), even $E_2$ disappears, leaving $K_5 \ot nS \to nS \ot E_1 \to K_5$ in addition to $P_{1, 0} \to K_1$ and $K_1 \to K_5$ to the list of regimes; of them only
 $P_{1, 0} \to K_1$ has realistic compactification. With further increase to $\mu = \mu_3$, the second ``touch'' happening -- see Fig.~\ref{D6_2}(e). And finally the situation for $\mu > \mu_3$ is presented in Fig.~\ref{D6_1}(f). The regimes there are: $P_{1, 0} \to K_1$, $K_5 \ot E_1 \to K_5$ (two different $K_5$) and $K_5 \ot nS \to nS \ot K_1$. One can see that of them only $P_{1, 0} \to K_1$ has viable compactification.

 This finalize our study of $D=6$ vacuum case. We see that the dynamics and its features are different from the previous cases, but the regimes with realistic compactification and their abundance is the same: all of the
 regimes require $\alpha > 0$ and there are two distinct regimes -- $P_{1, 0} \to E_{3+6}$ for $\mu < \mu_2$ (including $\beta < 0$) and $P_{1, 0} \to K_1$ for $\mu > \mu_2$.

 \subsection{$D=7$ case}

\begin{figure}
\centering
\includegraphics[width=0.96\textwidth, angle=0]{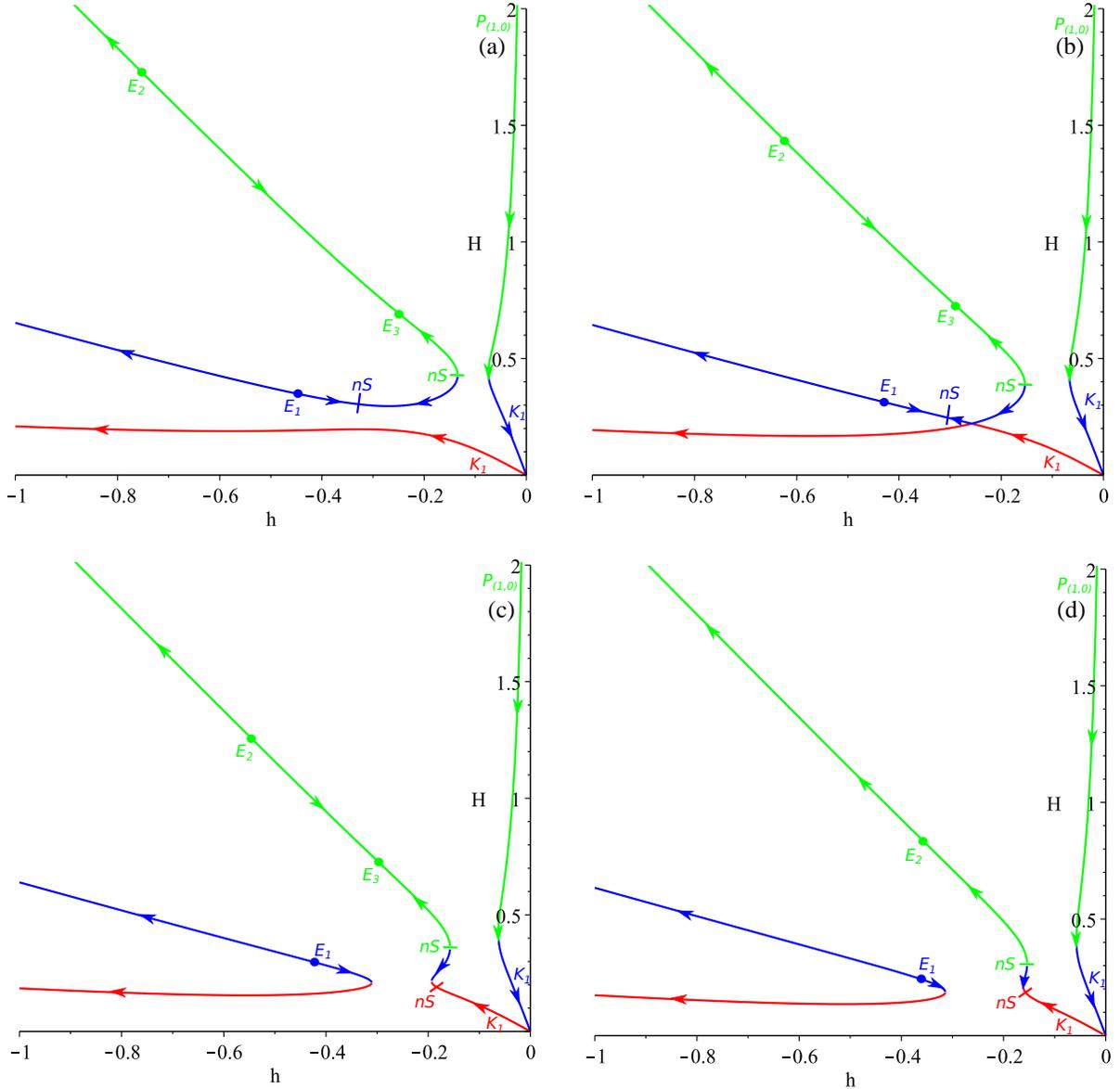}
\caption{Additional regimes which emerge in $D=7$ vacuum cubic Lovelock case, on $H(h)$ evolution curves; different colors correspond to three different branches $H_1$, $H_2$ and $H_3$;
panels layout is as follows: $\mu_3 > \mu > \mu_2$ on (a) panel, $\mu = \mu_3$ on (b) panel, $\mu_4 > \mu > \mu_3$ on (c) panel and $\mu = \mu_4$ on (d) panel
(see the text for more details).}\label{D7_1}
\end{figure}

The values for $\mu_i$ for $D=7$ are: $\mu_1 = 7/15 \approx 0.4667$, $\mu_2 \approx 0.2324$, $\mu_4 \approx 0.4084$ and $\mu_4 = 289/405 \approx 0.7136$. One can see that in the previous $D=6$ case we had
$\mu_3 > \mu_4 > \mu_2$ -- the disappearance of the exponential solutions along $H_3$ branch happened between the decouplings (at $\mu_{2, 3}$. Now, in $D=7$ case, we have $\mu_4 > \mu_3 > \mu_2$ --
the disappearance should happen after the second decoupling. So that the regimes up to $\alpha > 0$, $\beta > 0$, $\mu_4 > \mu > \mu_2$ (Fig.~\ref{D6_2}(b)) are exactly the same with the difference that now
Fig.~\ref{D6_2}(b) serves as a representative for $\mu_3 > \mu > \mu_2$ range. The remaining cases for $D=7$ are shown in Fig.~\ref{D7_1}. As always, different colors correspond to three different branches:
$H_1$ is red, $H_2$ is blue and $H_3$ is green, and the panels layout is the following: $\mu_3 > \mu > \mu_2$ in Fig.~\ref{D7_1}(a), $\mu = \mu_3$ in Fig.~\ref{D7_1}(b), $\mu_4 > \mu > \mu_3$ in Fig.~\ref{D7_1}(c)
and $\mu = \mu_4$ in Fig.~\ref{D7_1}(d); for $\mu > \mu_4$ the situation and the regimes are the same as in $\mu > \mu_3$ for $D=6$ and so are presented in Fig.~\ref{D6_1}(f).

Let us have a closer look on new regimes in this case. The first case, $\mu_3 > \mu > \mu_2$, is presented in Fig.~\ref{D7_1}(a) and is exactly the same as $\mu_4 > \mu > \mu_2$ from $D=6$, presented in Fig.~\ref{D6_2}(b)
(one can compare and verify that the regimes are the same). Then, the regime with realistic compactification is also the same -- it is $P_{1, 0} \to K_1$. The next case is $\mu = \mu_3$, presented in Fig.~\ref{D7_1}(b)
 -- the second ``detouch'' -- and the
regimes are the same. With further increase of $\mu$ to the $\mu_4 > \mu > \mu_3$ range, the situation is presented in Fig.~\ref{D7_1}(c) and the regimes are: $P_{1, 0} \to K_1$, $K_5 \ot E_1 \to K_5$ (with two different $K_5$)
and $K_5 \ot E_2 \to E_3 \ot nS \to nS \ot K_1$. One can see that the regimes along the last mentioned branch are the same as they are in the appropriate $\mu$ range for $D=6$. The realistic compactification in this case is
only $P_{1, 0} \to K_1$. The next case is $\mu = \mu_4$, presented in Fig.~\ref{D7_1}(d), and the changes are the same as they are in $D=6$ -- a pair of anisotropic exponential solutions $E_2$ and $E_3$ collapsed to a single
$E_2$ with $\sum H = 0$; the regimes are $K_5 \ot E_2 \ot nS \to nS \ot K_1$ and the only realistic compactification regime is $P_{1, 0} \to K_1$. Finally, for $\mu > \mu_4$ $E_2$ also disappears and the situation and the
regimes are identical to those in $\mu > \mu_3$ $D=6$ case, presented in Fig.~\ref{D6_1}(f).

This concludes our study of $D=7$ vacuum case. We see that the regimes demonstrate difference from $D=6$ case, but the realistic compactifications and their abundances remain the same: all
 regimes require $\alpha > 0$ and there are two of them -- $P_{1, 0} \to E_{3+7}$ for $\mu < \mu_2$ (including $\beta < 0$) and $P_{1, 0} \to K_1$ for $\mu > \mu_2$.

\subsection{General $D \geqslant 8$ case}

\begin{figure}
\centering
\includegraphics[width=0.7\textwidth, angle=0]{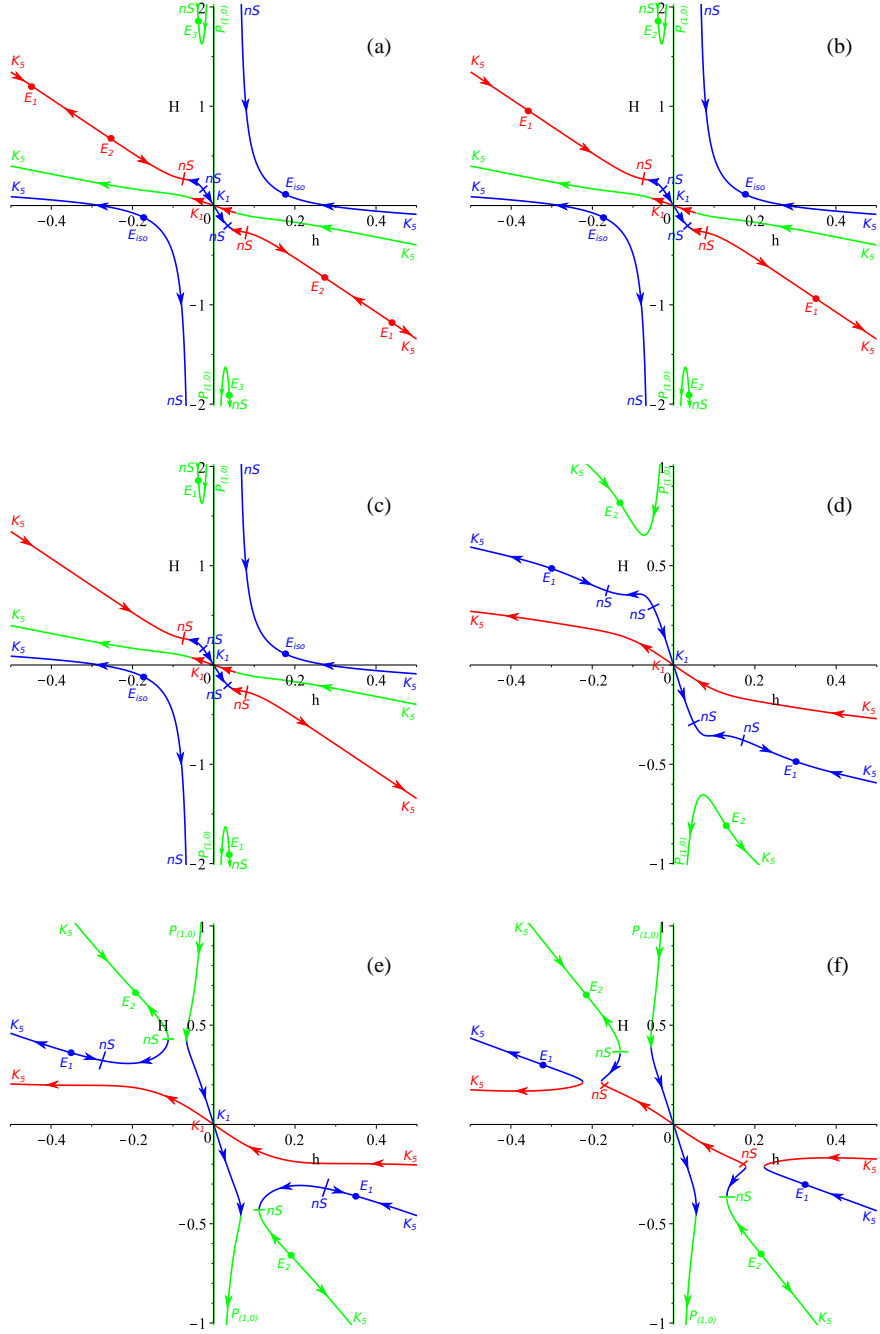}
\caption{Additional regimes which emerge in $D\geqslant 8$ vacuum cubic Lovelock case, on $H(h)$ evolution curves; different colors correspond to three different branches $H_1$, $H_2$ and $H_3$;
panels layout is as follows: $\alpha > 0$, $\beta < 0$, $\mu < \mu_4$ on (a) panel, $\alpha > 0$, $\beta < 0$, $\mu = \mu_4$ on (b) panel, $\alpha > 0$, $\beta < 0$, $\mu > \mu_4$ on (c) panel,
$\alpha > 0$, $\beta > 0$, $\mu < \mu_2$ on (d) panel, $\alpha > 0$, $\beta > 0$, $\mu_3 > \mu > \mu_2$ on (e) panel and $\alpha > 0$, $\beta > 0$, $\mu > \mu_3$ on (f) panel
(see the text for more details).}\label{D8_1}
\end{figure}

As we have learned previously, $\mu_4$ is the value for $\mu$ where the change of the number of anisotropic exponential solutions occurs. In all previous cases $D = 3 \div 7$ this value is positive, so that this change is happening in $\alpha > 0$, $\beta > 0$. But starting $D \geqslant 8$ it is negative, so that the change happening at $\alpha > 0$, $\beta < 0$. As a result, the regimes for $\alpha < 0$ are exactly the same as in $D =6,7$
(see Figs.~\ref{D6_1}(a)--(c)),
and there are no realistic compactifications. New regimes appear starting $\alpha > 0$, $\beta < 0$ are they are presented in Fig.~\ref{D8_1}. As usual, different colors correspond to three different branches:
$H_1$ is red, $H_2$ is blue and $H_3$ is green, and the panels layout is the following:  $\alpha > 0$, $\beta < 0$, $\mu < \mu_4$ is in Fig.~\ref{D8_1}(a), $\alpha > 0$, $\beta < 0$, $\mu = \mu_4$ is in Fig.~\ref{D8_1}(b), $\alpha > 0$, $\beta < 0$, $\mu > \mu_4$ is in Fig.~\ref{D8_1}(c),
$\alpha > 0$, $\beta > 0$, $\mu < \mu_2$ is in Fig.~\ref{D8_1}(d), $\alpha > 0$, $\beta > 0$, $\mu_3 > \mu > \mu_2$ is in Fig.~\ref{D8_1}(e) and $\alpha > 0$, $\beta > 0$, $\mu > \mu_3$ is in Fig.~\ref{D8_1}(f). We skipped
cases with exact $\mu = \mu_{2, 3}$ as the situation with exact value was described above. Let us have a closer look on individual panels.

In Fig.~\ref{D8_1}(a) we have presented $\alpha > 0$, $\beta < 0$, $\mu < \mu_4$ case. One can see that the behavior along hyperbola-like $H_2$, edge-shaped part of $H_3$ and $H_1-H_3$ physical branch is the same as in
$D=6$ case (compare with Fig.~\ref{D6_1}(d)), the changes are introduced to $H_1-H_2$ physical branch. One of the changes regards the number of the anisotropic exponential solutions and another -- the time direction of the evolution. One can see that ever since this branch (or its predecessor) appear in $D=4$ (see~\cite{my18b}) and in following cases $D=5$ (see Fig.~\ref{D5_1}(d)) and $D=6, 7$ (see Fig.~\ref{D6_1}(d)), $K_5$ always was
future attractor -- the regime was $E_1 \to K_5$. But in $D \geqslant 8$ it is reversed -- $K_5$ now is past attractor and can serve as a cosmological singularity, making $K_5  \to E_1$ successful compactification.
Apart from it, another realistic compactification regime is $P_{1, 0} \to E_2$ from edge-shaped part of $H_3$ branch. The other regimes along the physical branch under consideration include $E_2 \to E_1$, $E_2 \to nS$,
$nS \to nS$ and $K_1 \to nS$. To conclude, this is the first time in vacuum cubic Lovelock model when we see $K_5$ as an origin of viable compactification scheme. The next case is $\alpha > 0$, $\beta < 0$, $\mu = \mu_4$ and
it is presented in Fig.~\ref{D8_1}(b). Similar to the previously considered cases, at $\mu = \mu_2$ we have reduction of the anisotropic solutions -- $E_1$ and $E_2$ merge into a single constant volume $E_1$ solution; the
$K_5  \to E_1$ regime remains, as well as $P_{1, 0} \to E_2$. With further growth of $\mu$, to $\mu > \mu_4$, the situation is presented in Fig.~\ref{D8_1}(c). One can see that $E_1$ disappeared and so we have
$K_5 \to nS$ instead and only $P_{1, 0} \to E_1$ remains as a realistic compactification regime. So that, for $D \geqslant 8$ at $\alpha > 0$, $\beta < 0$, $\mu \leqslant \mu_4$ we have new realistic compactification
regime -- $K_5  \to E_1$.

The change of the time direction of one of the branches, described just above, affect $\alpha > 0$, $\beta > 0$ as well -- $K_5$ on $H_3$ branch, which is future asymptote in $D=4$ (see~\cite{my18b}),
$D=5$ (see Figs.~\ref{D5_1}(e, f)), $D=6$ (see Figs.~\ref{D6_1}(e, f)) and $D=7$ (see Figs.~\ref{D7_1}), now in $D \geqslant 8$ it is a past asymptote, in analogue with $\alpha > 0$, $\beta < 0$ case described above.
So that, similar to the previous case, we have $K_5 \to E_2$ realistic compactification, and it is present for all $\mu > 0$ (see Figs.~\ref{D8_1}(d)--(f)). Another realistic compactification regime is
$P_{1, 0} \to E_2$ for $\mu < \mu_2$ (see Fig.~\ref{D8_1}(d)) and $P_{1, 0} \to K_1$ for $\mu > \mu_2$  (see Figs.~\ref{D8_1}(e), (f)). The details and other regimes (without compactifications)
are quite similar to the analogues in previous case so we skip their consideration.

To conclude, general $D \geqslant 8$ bring us new realistic compactification regime -- \mbox{$K_5 \to E_{3+D}$} -- the only regime with realistic compactification which originates from $K_5$. The regime appear for $\alpha > 0$,
$\beta < 0$, $\mu \leqslant \mu_4$ and the entire $\alpha > 0$, $\beta > 0$.

\section{Discussions}

In the current paper we have analyzed the cosmological dynamics of the cubic Lovelock gravity, with Einstein-Hilbert and Gauss-Bonnet terms present as well. We have chosen a setup with a topology being a product of two
isotropic subspaces -- three-dimensional, representing our Universe, and $D$-dimensional, representing extra dimensions. Both subspaces are flat, which simplifies our equations of motion and makes it possible to analyze them
analytically. Current paper is a direct continuation of~\cite{my18b} where we considered low-$D$ ($D=3,4$) cases and in current paper we extend the analysis on the remaining
high-$D$ cases.
In a sense, it is also a logical continuation of~\cite{my16a, my16b, my17a, my18a}, where we considered the same problem but in EGB gravity -- vacuum case in~\cite{my16a, my18a} and $\Lambda$-term case in~\cite{my16b, my17a, my18a}. In~\cite{my18a} we reviewed all the results for EGB from~\cite{my16a, my16b, my17a} and changed the visualization of the regimes -- in the original papers~\cite{my16a, my16b, my17a} we use tables to list of all the regimes, and this way sometimes is not easy to read. On the contrary, in~\cite{my18a} we put all
the regimes on $H(h)$ curves and added arrows to demonstrate $t\to\infty$ directed evolution. In the current paper we decided to keep visualization from~\cite{my18a}.

First of all, let us summarize the results, as they are scattered over mini-conclusions in each particular sections here and in~\cite{my18b}. The fist case is $D=3$, presented
in~\cite{my18b}
and it has interesting feature -- since the
equations of motion are cubic in both $H$ and $h$, there could be up to three branches of the solutions. On the other hand, it is the lowest possible dimension for cubic Lovelock
gravity, so there are no Kasner solutions (see~\cite{PT}). Then the only possibility is what we call ``generalized Taub'' solution -- the situation when the expansion in each direction
is characterized by Kasner exponent $p_i = -H_i^2/\dot H_i$ equal to either 1 or 0; so that for our topology it is either $P_{1, 0}$ ($p_H = 1, p_h = 0$ -- expansion of the three-dimensional subspace and ``static'' extra dimensions) or $P_{0, 1}$ ($p_H = 0, p_h = 1$ -- expansion of the extra-dimensional subspace and ``static'' three dimensions). Then
the remaining branches -- which cannot be connected to either $P_{1, 0}$ or $P_{0, 1}$, form closed evolution curves for ($\alpha < 0, \beta < 0$) and
($\alpha > 0, \beta > 0$); for ($\alpha < 0, \beta > 0$) and ($\alpha > 0, \beta < 0$) they encounter nonstandard singularities (see~\cite{my18b} for details).
The realistic compactification regimes are $P_{1, 0}/P_{0, 1} \to E_{3+3}$ for ($\alpha > 0, \beta < 0$) and
 $P_{1, 0}/P_{0, 1} \to K_1$ for ($\alpha > 0, \beta > 0$); let us note that both of the regimes exist only for $\alpha > 0$.

The $D=4$ case has one cubic Lovelock Kasner solution $K_5$ but it is still not enough for all branches, so we still nonstandard singularities at ($\alpha < 0, \beta > 0$) and
($\alpha > 0, \beta < 0$) while for ($\alpha < 0, \beta < 0$) and ($\alpha > 0, \beta > 0$) the evolution curves have complicated shapes.
In $D=4$ we still have $P_{1, 0}$ regime, but not $P_{0, 1}$, and some of the nonstandard singularities have power-law behavior
and so designated as $nS/P$. Unlike $D=3$, where the regimes within the fine structure existed on an isolated $H(h)$ curve, in $D=4$ they are located on
one of the physical branches connected with $K_1$ (see~\cite{my18b} for details).
The realistic compactification regimes are $P_{1, 0} \to E_{3+4}$ for $\alpha > 0$, $\mu < \mu_1$ (including entire $\beta < 0$) and $P_{1, 0} \to K_1$, for $\alpha > 0$, $\mu > \mu_1$ -- exactly the same as in $D=3$ case, and again both of the regimes exist only for $\alpha > 0$.

In $D=5$ (see Figs.~\ref{D5_1} and~\ref{D5_2}) we have two different $K_5$ and a ``return'' of $P_{0, 1}$, and the general evolution curves and the regimes resemble previous cases;
the same is true for the realistic compactification regimes -- the same $P_{1, 0} \to E_{3+5}$ for $\alpha > 0$, $\mu < \mu_1$ (including entire $\beta < 0$) and $P_{1, 0} \to K_1$, for $\alpha > 0$, $\mu > \mu_1$. So that, we can conclude that in $D=3, 4, 5$ the dynamics, with the exception of some features, is generally the same, so as the realistic compactification
regimes.

Formally the equations of motion keep the functional form in all $D \geqslant 6$ cases (meaning that no new terms appear, how it was in the lower dimensions), but our analysis suggests
that the details of the dynamics -- similar to the differences between $D=3, 4, 5$ -- are different in $D=6, 7$ and $D \geqslant 8$ cases, so that we considered them separately.
With a minor details, the dynamics of $D=6$ (see Fig.~\ref{D6_1}) and $D=7$ (see Fig.~\ref{D7_1}) do not differ much from the previous cases, and the realistic compactification
regimes are the same. The only difference is absence of the ``fine-structure'' of anisotropic exponential solutions, but instead we received the ``fine-structure'' of the $H(h)$ curves.
On the other hand, general $D \geqslant 8$ case brought us a whole new regime with realistic compactification -- $K_5 \to E_{3+D}$ -- a regime which originates from ``normal'' Kasner regime, instead of $P_{1, 0}$. This regime exist in two domains: $\alpha > 0$, $\beta < 0$, $\mu \leqslant \mu_4$ and entire $\alpha > 0$, $\beta > 0$. The regimes $P_{1, 0} \to E_1$ and
$P_{1, 0} \to K_1$ are also present in the general $D \geqslant 8$ case.

To conclude the situation with the realistic compactifications, for all $D \geqslant 3$ there are
$P_{1, 0} \to E_{3+D}$ regime for $\alpha > 0$, $\mu < \mu_1$ (including entire $\beta < 0$) and $P_{1, 0} \to K_1$ regime for $\alpha > 0$, $\mu > \mu_1$. For $D \geqslant 8$ there is
additional regime $K_5 \to E_{3+D}$ which exists in two domains: $\alpha > 0$, $\beta < 0$, $\mu \leqslant \mu_4$ and entire $\alpha > 0$, $\beta > 0$. Let us note that for
$D \geqslant 8$ and $\alpha > 0$, $\beta < 0$, $\mu < \mu_4$
 there are two realistic compactification regimes which exist at the same time and have two different anisotropic exponential solutions as a
future asymptotes -- $K_5 \to E_1$ along $H_2$ branch and $P_{1, 0} \to E_3$ on edge-shaped part of $H_3$ branch (see Fig.~\ref{D8_1}(a)). For $D \geqslant 8$ and $\alpha > 0$,
$\beta > 0$, $\mu < \mu_2$ there are two realistic compactification regimes but they lead to the same anisotropic exponential solution -- $K_5 \to E_2$ and $P_{1, 0} \to E_2$, both
on the same $H_3$ branch (see Fig.~\ref{D8_1}(d)).

The above-mentioned ``generalized Taub'' solution deserves additional comments. Formally it fits the description of the ``generalized Milne'' solution -- the second branch of the power-law solutions
in Lovelock gravity (see~\cite{prd09} for details), but only formally -- it fits only because it is degenerative. As we demonstrated in~\cite{PT}, strict ``generalized Milne''
cannot exist in pure highest-order Lovelock gravity, as it leads to degeneracy in the equations of motion. But if additional (lower-order) Lovelock contributions are involved, it this branch of solutions could be restored, but it was never demonstrated before. So that on the particular example of $[3+D]$ spatial splitting we demonstrated this possibility. Still,
a little is known about this regime and it deserves additional investigation in the separate papers.

When we consider this ``generalized Taub'' solution as a past asymptote -- and this is the case for all possible realistic compactification models in $D=3 \div 7$ -- it feels
unnatural. Indeed -- the $P_{1, 0}$ regime imply $H \to \infty$ and $h \to 0$ as $t \to 0$ (by ``0'' we mean here initial cosmological singularity), so that we initially have ``burst''-like
expansion of three-dimensional subspace while the extra-dimensional subspace is almost static. In addition to the feeling of unnaturalness, it is a question if this regime could be
reached from totally anisotropic space, in a manner it was done in~\cite{PT2017} for Kasner regimes in EGB case. So that it gives additional reason to deeply investigate this regime and we are going
to do it in the nearest time.

Similar to the results of~\cite{my18b}, the results of this paper suggest that the variety and abundance of the regimes is closer to $\Lambda$-term EGB, rather then to the vacuum EGB models. The reasons for that are not exactly clear, but we suspect that number of the free
parameters plays a role here. Indeed, for vacuum EGB model there is only one parameter -- $\alpha$, Gauss-Bonnet coupling, while for $\Lambda$-term EGB and vacuum cubic Lovelock there are two -- $\alpha$ and $\Lambda$ for
the former and $\alpha$ and $\beta$ (cubic Lovelock coupling) for the latter. So that we expect that the dynamics of the $\Lambda$-term cubic Lovelock gravity to be even more interesting and
we are going to consider this case shortly.

\section{Conclusions}

This concludes our study of the cosmological models in vacuum cubic Lovelock gravity. We have found that in all $D \geqslant 3$ there are compactification regimes of two kinds,
the first of them originate from ``generalized Taub'' solution; for the future asymptote we have either Kasner regime or anisotropic exponential solution.
In $D \geqslant 8$ there appears another compactification scheme which originates from high-energy Kasner regime and has anisotropic exponential solution as future
asymptote. So that for $D \geqslant 8$ and some parameter the two of them coexist on different branches - the situation we never had in EGB gravity.

In addition to the regimes with successful compactification, we described and plotted on $H(h)$ curves all possible transitions for all initial conditions and all structurally different cases.
The variety and abundance of the regimes exceed even $\Lambda$-term EGB case, featuring transition between two anisotropic exponential solutions and transition between two different
``generalized Taub'' solutions.

There are two interesting observations which require additional investigation, as both are quite unexpected. First of them
 is that all of the realistic compactification regimes have $\alpha > 0$ requirement. This is unexpected, as in both vacuum and $\Lambda$-term EGB cases we have viable
 compactifications for both signs of $\alpha$. We can note that for the $\Lambda$-term case the joint analysis of our cosmological bounds and those coming from AdS/CFT and other considerations allows
 us to conclude $\alpha > 0$ (see~\cite{my16b, my18a}), but for that we involved external (to our analysis) results. On the contrary, in the current case without any external bounds we already have
 realistic compactification only for $\alpha > 0$.

 The second observation is that there is no $K_5 \to K_1$ transition with realistic compactification.  In EGB vacuum case~\cite{my16a, my18a} we have the transitions of this kind, so we expected that in the higher-order Lovelock gravity they would also appear, but our investigation reveals that they do not. There is $K_5 \to K_1$ transition, but with contracting
 three and expanding extra dimensions, so it formally exist, but with no realistic compactification. As both of these observations are unexpected and are in disagreement with what we have
 learned from study of EGB case, this is a good direction for further improvement of our understanding of Lovelock gravity.

In fact, the cubic (and higher-order) Lovelock gravities are studied much less then the Gauss-Bonnet gravity (which is quadratic Lovelock gravity) -- apart from the above-mentioned
papers, we studied some properties of the power-law~\cite{prd09} and exponential~\cite{CST2} solutions and studied the stability of the latter~\cite{my15}. Additionally,
some properties of models with spatial curvature are studied in~\cite{CGT18}. Our results suggest that there are some
interesting features making the dynamics of the cubic Lovelock gravity different from EGB case, which increase the significance of the results and stimulate further investigation
of cubic and higher-order Lovelock cosmologies.

\end{document}